\documentclass[a4paper,11pt]{article}

\usepackage{jheppub}
\usepackage{xpatch}

\makeatletter
\xpatchcmd{\maketitle}
  {\noindent{\small\scshape\@fpheader}\@preprint\par}
  {\noindent{\small\scshape\hphantom{Prepared for submission to JHEP}}\@preprint\par}
  {}{}
\makeatother

\usepackage[T1]{fontenc}
\usepackage[utf8]{inputenc} 
\usepackage[english]{babel}
\usepackage{csquotes}
\usepackage{enumitem}

\usepackage{mathtools} 
\usepackage{mathrsfs}
\usepackage{bm}
\usepackage{amssymb,amsmath}

\usepackage{graphicx}
\usepackage{booktabs}
\usepackage{multirow}
\usepackage{array} 
\newcolumntype{C}[1]{>{\centering\let\newline\\\arraybackslash\hspace{0pt}}m{#1}}
\newcolumntype{L}[1]{>{\let\newline\\\arraybackslash\hspace{0pt}}m{#1}}
\usepackage{placeins}
\usepackage{subcaption}

\usepackage[dvipsnames]{xcolor}
\usepackage{xspace}
\usepackage{enumitem}
\usepackage{changepage}
\usepackage{marginnote}
\usepackage{siunitx}

\usepackage[colorlinks=true,citecolor=blue,linkcolor=blue]{hyperref}
\usepackage[capitalize]{cleveref}
\crefname{appendix}{Appendix}{Appendices}
\Crefname{appendix}{Appendix}{Appendices}

\graphicspath{{../}}

\definecolor{Morange}{rgb}{0.880722,0.611041,0.142051}
\definecolor{Mgreen}{rgb}{0.560181,0.691569,0.194885}
\definecolor{Mred}{rgb}{0.922526,0.385626,0.209179}
\definecolor{dfa}{rgb}{0., 0., 0.}
\definecolor{dfb}{rgb}{0.8862745098039215, 0.2901960784313725, 0.2}
\definecolor{dfc}{rgb}{0.20392156862745098, 0.5411764705882353, 0.7411764705882353}
\definecolor{dfd}{rgb}{0.596078431372549, 0.5568627450980392, 0.8352941176470589}
\definecolor{dfe}{rgb}{0.4666666666666667, 0.4666666666666667, 0.4666666666666667}
\definecolor{dff}{rgb}{0.984313725490196, 0.7568627450980392, 0.3686274509803922}
\definecolor{dfg}{rgb}{0.5568627450980392, 0.7294117647058823, 0.2588235294117647}
\definecolor{dfh}{rgb}{1., 0.7098039215686275, 0.7215686274509804}

\definecolor{cborange}{HTML}{e69f00}
\definecolor{cbgreen}{HTML}{009e73}
\definecolor{cbyellow}{HTML}{f1dd42}
\definecolor{cblblue}{HTML}{56b4e9}
\definecolor{cbblue}{HTML}{0072b2}
\definecolor{defgrey}{HTML}{9f9f9f}
\definecolor{defgreen}{HTML}{8eba42}

\newcommand{\gray}[1]{{\color{defgrey}  #1}}
\usepackage{xpatch}
\makeatletter
\patchcmd{\@ssect@ltx}
    {\addcontentsline{toc}{#1}{\protect\numberline{}#8}}
    {}
    {}
    {}

\DeclareSIUnit\year{yr}
\DeclareSIUnit\years{yrs}

\makeatletter
\providecommand*{\diff}%
	{\@ifnextchar^{\DIfF}{\DIfF^{}}}
\def\DIfF^#1{%
	\mathop{\mathrm{\mathstrut d}}%
		\nolimits^{#1}\gobblespace}
\def\gobblespace{%
	\futurelet\diffarg\opspace}
\def\opspace{%
	\let\DiffSpace\!%
	\ifx\diffarg(%
		\let\DiffSpace\relax
	\else
		\ifx\diffarg[%
			\let\DiffSpace\relax
		\else
  			\ifx\diffarg\{%
				\let\DiffSpace\relax
			\fi\fi\fi\DiffSpace}

\newcommand{\Lumgeom}{\mathscr{L}_\text{geom}}

\usepackage[nolist]{acronym}
\begin{acronym}
	\acro{SM}{Standard Model}
	\acro{BSM}{beyond the Standard Model}
	\acro{COM}{center-of-mass}
	\acro{QED}{Quantum Electrodynamics}
	\acro{EFT}{Effective Field Theory}
	\acro{RHS}{right hand side}
	\acro{LHS}{left hand side}
	\acro{EM}{Electromagnetism}
	\acro{SNR}{signal to noise ratio}
	\acro{CL}{confidence level}
	\acro{VEV}{vacuum expectation value}
	\acro{EWSB}{Electroweak Symmetry Breaking}
	\acro{PDF}{parton distribution function}
	\acro{LHC}{Large Hadron Collider}
    \acro{HL-LHC}{High Luminosity LHC}
    \acro{BR}{branching ratio}
    \acro{DOF}{degrees of freedom}
    \acro{SUSY}{Supersymmetry}
    \acro{MSSM}{minimally supersymmetric Standard Model}
    \acro{HSCP}{heavy stable charged particle}
    \acro{pCM}{parton-center-of-mass energy}
    \acro{RF}{radio frequency}
    \acro{LWFA}{laser-wakefield acceleration}
    \acro{PWFA}{particle-beam wakefield acceleration}
    \acro{IP}{interaction point}
    \acro{EW}{electroweak}
    \acro{VBF}{vector-boson fusion} 
    \acro{WBF}{$W$-boson fusion}
    \acro{ZBF}{$Z$-boson fusion}
\end{acronym}

\newcommand\myshade{80}
\colorlet{mylinkcolor}{ForestGreen}
\colorlet{mycitecolor}{Red}
\colorlet{myurlcolor}{violet}

\hypersetup{
  linkcolor  = mylinkcolor!\myshade!black,
  citecolor  = mycitecolor!\myshade!black,
  urlcolor   = myurlcolor!\myshade!black,
  colorlinks = true
}

\title{Searches for electroweak singlets at future plasma wakefield colliders}

\preprint{
{\scriptsize
  \begin{tabular}[t]{@{}r@{}}
    CERN-TH-2026-279\\
    MIT-CTP/6127\\
    SISSA 23/2026/FISI
  \end{tabular}
  }
}

\author[a]{So Chigusa,}
\author[b,c]{Simon Knapen,}
\author[d,e]{Toby Opferkuch,}
\author[a,b,c]{Inbar Savoray,}
\author[b,c,h]{Christiane Scherb,}
\author[f,g]{and Weishuang Linda Xu}

\affiliation[a]{Center for Theoretical Physics -- a Leinweber Institute, Massachusetts Institute of Technology,\\
77 Massachusetts Avenue, Cambridge, MA 02139, USA}
\affiliation[b]{Theoretical Physics Group, Lawrence Berkeley National Laboratory, Berkeley, CA 94720, USA}
\affiliation[c]{Leinweber Institute for Theoretical Physics, Department of Physics, University of California, \mbox{Berkeley}, CA 94720, USA}
\affiliation[d]{SISSA International School for Advanced Studies, Via Bonomea 265, 34136, Trieste, Italy}
\affiliation[e]{INFN, Sezione di Trieste, Via Bonomea 265, 34136, Trieste, Italy}
\affiliation[f]{Kavli Institute for Particle Astrophysics \& Cosmology, Stanford University, \\Stanford, CA 94305, USA}
\affiliation[g]{Particle Theory Group, SLAC National Accelerator Laboratory, Stanford, CA 94305, USA}
\affiliation[h]{Theoretical Physics Department, CERN, 1 Esplanade des Particules, CH-1211 Geneva 23, Switzerland}

\emailAdd{schigusa@mit.edu}
\emailAdd{smknapen@lbl.gov}
\emailAdd{toby.opferkuch@sissa.it}
\emailAdd{inbarsav@mit.edu}
\emailAdd{christiane.scherb@cern.ch}
\emailAdd{wlxu@stanford.edu}

\abstract{
We study the sensitivity of future \SI{10}{\TeV} plasma wakefield colliders to a real electroweak singlet scalar $S$ coupled to the Standard Model through the Higgs portal.
Using luminosity spectra obtained from particle-in-cell simulations, we compare $e^+e^-$ and $e^-e^-$ configurations with both flat and round beam profiles, as well as a $\gamma\gamma$ collider based on Compton backscattering of laser photons off the $e^-e^-$ beams.
The $e^-e^-$ and $\gamma\gamma$ options avoid the open challenge of accelerating positrons in plasma.
Without a positron beam, however, the vector boson fusion channels $e^+e^-\to\nu\bar{\nu}S$ and $e^+e^-\to\nu\bar{\nu}SS$, which dominate at $e^+e^-$ colliders, can only proceed through secondary positrons and are heavily suppressed.
At a $\gamma\gamma$ collider, this loss in signal rate is partially compensated by additional photon-initiated processes.
While the $e^-e^-$ configurations are not as competitive, both the $e^+e^-$ and $\gamma\gamma$ wakefield options have a reach qualitatively similar to that of a \SI{10}{\TeV} muon collider and the FCC-hh.
Between the two, the $\gamma\gamma$ option gives the stronger sensitivity if the achievable geometric luminosities of $e^+e^-$ colliders are more than a factor of $\sim$3 lower than that of the $\gamma\gamma$ collider.  }
\date{\today}

\begin{document}
\maketitle

\section{Introduction}
\label{sec:intro}

Following the LHC's success, the particle physics community is exploring complementary paths toward the next major energy leap.
Advancing the energy frontier by an order of magnitude means reaching parton center-of-mass energies of about \SI{10}{\TeV}, which would require a \SI{10}{\TeV} electron or muon collider, or a proton collider at roughly \SI{100}{\TeV}.
The Future Circular Collider (FCC) ultimately targets \SI{85}{\TeV} through an ambitious scaling of proven concepts~\cite{FCC:2025uan}, while muon and plasma wakefield colliders are being pursued as less mature but potentially transformative technologies for multi-TeV lepton collisions \cite{Gessner:2025acq,Verra:2026rvs}.

Facility size and cost are central challenges, especially for high-energy electron colliders.
Wakefield accelerators aim to address this by exploiting collective plasma oscillations to produce multi-GV/m accelerating gradients, orders of magnitude above the $\mathcal{O}(100)\,\mathrm{MV/m}$ scale attainable with RF cavities \cite{Tajima:1979,Chen:1985,Esarey:2009,Adli:2019,CLICdp:2018cto}.
An intense driver, either a laser pulse or a high-current particle bunch, perturbs the plasma and generates a trailing longitudinal electric field that efficiently accelerates a second bunch of charged particles following the driver, known as the witness bunch.
These large gradients could in principle allow a \SI{10}{\TeV} electron collider to fit within the footprint of a conventional linear collider such as the ILC \cite{Gessner:2025acq,Roser:2022sht}.
While plasma wakefield acceleration has been experimentally demonstrated \cite{Leemans:2006dx,Blumenfeld:2007ph,Litos:2014yqa}, further R\&D is needed to arrive at a compact, high-energy implementation with sufficiently high luminosity.
In particular, to realize this vision, challenges such as the preservation of beam quality (sub-$\mu$m emittance, percent-level energy spread), multi-stage synchronization, high repetition rate, and control of beam--beam effects at the \ac{IP} must all be addressed \cite{Esarey:2009,Adli:2019,Schroeder:2022xdu,Diederichs:2024zir}.
Recently, the \SI{10}{\TeV} \ac{pCM} Wakefield Collider Design study was launched \cite{Gessner:2025acq} to chart this R\&D path, aiming to deliver an end-to-end design concept with self-consistent parameters and an overall cost estimate.
Achieving these goals will require selecting among competing technologies to focus future R\&D efforts.

Two accelerator challenges are particularly important for the physics potential of these facilities.
First, the lack of high-efficiency concepts for positron acceleration in plasma wakefields as well as intense enough positron sources severely complicates the design of the positron arm of the wakefield collider \cite{PhysRevE.64.045501,Hogan:2003bs}.
Despite substantial R\&D progress, the most promising theoretical concepts achieve a predicted luminosity per unit beam power roughly an order of magnitude below that achievable with \ac{RF} technology \cite{PhysRevAccelBeams.27.034801}.
This challenge has motivated the consideration of $e^-e^-$ and $\gamma\gamma$ colliders as alternatives to the more challenging $e^+e^-$ option.
In the $\gamma\gamma$ case, photons from a pair of intense lasers are Compton-backscattered off the electron beams, producing high-energy photon beams \cite{Ginzburg:1981vm}.
Some of the backscattered photons also collide with laser photons and produce high-energy $e^+e^-$ pairs through the Breit--Wheeler process \cite{Barklow:2023ess}. 

Second, at a \SI{10}{\TeV} electron wakefield collider, the intense electromagnetic fields at the \ac{IP} induce strong beam--beam interactions that disrupt the beams, give the luminosity spectrum a long low-energy tail through beamstrahlung, and generate large fluxes of secondary $\gamma$ and $e^\pm$ particles~\cite{YokoyaChen:1992}.
These secondaries strongly affect detector design and particle production rates for both signal and background processes, and it remains an open question to what extent these effects must be mitigated, or can instead be exploited, in a high-energy discovery machine (see e.g.~\cite{Barklow:2023iav}).
Specifically, the intense electromagnetic fields at the \ac{IP} cause beamstrahlung: hard photon radiation that broadens the electron energy spectrum and shifts it toward lower energies \cite{Esberg:2014zia,RevModPhys.94.045001}.\footnote{Recently, a set of analytic approximations has been developed tailored to the beam--beam interactions in this regime \cite{He:2025jmz}.} The beamstrahlung rate is minimized for flat beams, which have asymmetric dimensions in the transverse direction \cite{Schroeder_2022}.
However, nonlinear transverse wakefields in plasma accelerators tend to mix the transverse emittances, severely degrading the quality of flat beams \cite{Diederichs:2024zir}.

Importantly, these secondary particles not only produce backgrounds but also provide sizable $e^\pm\gamma$, $\gamma\gamma$, and secondary $e^+e^-$ luminosities.
Taken together, these accelerator challenges and beam-induced opportunities motivate a comparison of the physics reach of $e^+e^-$ and $e^-e^-$ colliders, in both round- and flat-beam configurations, and of a $\gamma\gamma$ collider based on round electron beams.
Such a comparison will allow the community to assess \emph{whether the $e^+$ acceleration and/or the flat-beam delivery challenges can be bypassed without compromising our physics goals.}
An affirmative answer to either question, or to both, could greatly accelerate the path to an energy-frontier wakefield accelerator.
To this end, a joint effort has brought together expertise in accelerator physics, experimental particle physics, and particle theory.
This includes work that addressed accelerator optimization \cite{Bulanov:2026nqa}, modeling of beam--beam interactions \cite{Formenti:2026jun}, and forthcoming work on the constraints those interactions place on a realistic detector design \cite{detectorpaper}. 
On the theory side, this program requires evaluating the discovery potential of each collider configuration across a broad set of physics targets. Such studies have already been completed for pair production of charged particles \cite{Chigusa:2025azs}, heavy vector resonances \cite{Cipressi:2026wtb} and precision measurements of the Higgs potential \cite{Fraser:2026ngs}. In all cases, the $\gamma\gamma$ collider emerged as a promising alternative to the $e^+e^-$ colliders. A similar study for sources of lepton flavor violation is currently underway \cite{LVF}.

In this paper, we study the discovery potential for an electroweak-singlet scalar coupled to the Standard Model through the Higgs portal, complementing our prior work on particles charged under the electroweak force \cite{Chigusa:2025azs}.
Scalar singlets are minimal Higgs-portal extensions that appear in a range of settings, including dark matter models~\cite{McDonald:1993ex,Burgess:2000yq}, scenarios with a first-order electroweak phase transition ~\cite{Profumo:2014opa,Curtin:2014jma,Noble:2007kk,Kotwal:2016tex}, neutral-naturalness constructions~\cite{Chacko:2005pe}, and supersymmetric extensions~\cite{Ellwanger:2009dp}.
For the Higgs-portal benchmark considered here, single production of an electroweak singlet at multi-TeV energies is dominated by \ac{VBF}, rather than the direct annihilation channels available for electroweakly charged states.
At an $e^+e^-$ collider, the leading charged-current channel, $e^+e^-\to\nu_e\bar\nu_e S$, can proceed directly, whereas at $e^-e^-$ and $\gamma\gamma$ colliders the same channel requires secondary positrons.
Although the $e^-e^-\to e^-e^- S$ channel is accessible through neutral-current $Z$-fusion, its cross section is about an order of magnitude below that of $W$-boson fusion at the same collision energy, because the electron--$Z$ couplings in the \ac{SM} happen to be small. 
The gap in event yields between the $e^+e^-$ and $e^-e^-$ machines at equal geometric luminosity is larger still, nearly two orders of magnitude for round beams (see \cref{sec:singletsignalprod}).
This poses an apparent disadvantage for the $e^-e^-$ and $\gamma\gamma$ configurations.
We show, however, that the additional $W$-fusion topologies $\gamma\gamma\to W^+W^-S$ and $e^-\gamma\to \nu_e W^- S$ largely compensate for the loss of the primary $e^+e^-$ charged-current channel, particularly for a $\gamma\gamma$ machine.
We also study pair production, which is controlled by the quartic Higgs portal coupling and can remain sizable when single production is suppressed by a small singlet--Higgs mixing.
We investigate both prompt hadronic and displaced decays, which provide complementary sensitivity in this regime.

This paper is organized as follows:
We define and motivate our benchmark model in \cref{sec:singletmodel}. We then analyze single and pair production modes in \cref{sec:singprod} and \cref{sec:singletpairprod}, respectively. We conclude in \cref{sec:conclusions} and reserve additional analysis details and a brief discussion of some alternative strategies to the appendices.

\section{Benchmark model\label{sec:singletmodel}}

In this work we will discuss the search for models where the Standard Model is augmented with one heavy electroweak singlet scalar, whose interactions with the \ac{SM} occur exclusively via the Higgs sector.
From a theoretical point of view these singlets are a feature of several well-motivated theories such as the twin Higgs model~\cite{Chacko:2005pe} and the NMSSM~\cite{Ellwanger:2009dp}.
If the scalar is stable, it may be a viable candidate for dark matter~\cite{McDonald:1993ex,Burgess:2000yq}.
If it substantially modifies the Higgs potential it may catalyze a first order phase transition~\cite{Profumo:2014opa,Curtin:2014jma,Noble:2007kk,Kotwal:2016tex}. 
From a phenomenological perspective, this scenario is appealingly minimal and its discovery potential is intimately related to our ability to characterize the Higgs sector.
The heavy singlet scalar has been studied as a physics target for other candidate future colliders, including lepton~\cite{Buttazzo:2018qqp,Ruhdorfer:2019utl,Aboudonia:2024frg}, hadron~\cite{Kotwal:2016tex,Curtin:2014jma} and photon \cite{Garcia-Abenza:2020xkk}.
In a certain sense, the ability of a wakefield collider to discover this class of models is reflective of its ability to deliver on two core demands of a next-generation high energy experiment: the search for undiscovered particles, and the exploration of the Higgs potential. 

The renormalizable terms involving a real singlet $S_0$ can be written as
\begin{equation}\label{eq:lagdefinition}
   \mathcal{L} \supset \frac{1}{2}(\partial_\mu S_0)^2-\frac{1}{2}m_{S,0}^2 S_0^2-a_{HS}S_0|H|^2-\frac{\lambda_{HS}}{2}S_0^2|H|^2-\frac{a_S}{3}S_0^3-\lambda_H|H|^4-\frac{\lambda_S}{4}S_0^4\,,\\
\end{equation}
where $H$ is the \ac{SM} Higgs doublet 
\begin{equation}
 H = \begin{pmatrix}
\pi^+ \\
\frac{v + h_0 + i \pi_0}{\sqrt{2}}
\end{pmatrix}\,.
\end{equation} 
Here $v$ and $h_0$ are the expectation value and radial mode respectively; the $\pi$ are the Goldstone modes that make up the longitudinal components of the $W$ and $Z$ bosons.
Note we have implicitly shifted the singlet field so that $\langle S_0\rangle=0$ and invoked the tadpole condition to eliminate the linear term.
 
 The $h_0$ and $S_0$ fields mix, and we will denote the light and heavy mass eigenstates with $h$ and $S$ respectively, with masses $m_h$ and $m_S$. 
$S$ then inherits couplings to the \ac{SM} particle content via mixing with the \ac{SM} Higgs.
This mixing is parametrized by the rotation angle $\gamma$ between the interaction and mass bases,
\begin{align}
    h = h_0\cos\gamma + S_0 \sin\gamma\,, \quad\quad S = S_0\cos\gamma-h_0\sin\gamma\,, 
\end{align}
where the mixing angle can be written explicitly as
\begin{align}\label{eq:mixingangle}
\sin\gamma \simeq  -a_{HS}  \frac{v}{m_S^2 - m_h^2}\, .
\end{align}
This mixing enables the singlet to couple directly to the gauge sector of the \ac{SM}.
For scalar and gauge couplings defined as
\begin{align}
    \mathcal{L} \supset& - \frac{1}{2} g_{hhS} h^2 S - \frac{1}{2} g_{hSS} h S^2 - \frac{1}{4} g_{hhSS} h^2 S^2 
    + g_{WWS} S W_\mu^+ W^{-\mu} + \frac{1}{2} g_{ZZS} S Z_\mu Z^\mu \notag \\
    & \phantom{- \frac{1}{2} g_{hhS} h^2 S - \frac{1}{2} g_{hSS} h S^2 - \frac{1}{4} }+ \frac{1}{2} g_{WWSS} S^2 W_\mu^ + W^{-\mu} + \frac{1}{4} g_{ZZSS} S^2 Z_\mu Z^\mu\,,
\end{align}
these couplings are given at tree level by
\begin{align}
    g_{hhS} &= -\frac{s_\gamma c^2_\gamma}{v}(m_S^2+2m_h^2)
      +\lambda_{HS}vs_\gamma(2 c^2_\gamma-s^2_\gamma)
      +2a_S c_\gamma s^2_\gamma\,, \\
    g_{hSS} &=\frac{ c_\gamma s^2_\gamma}{v}(m_h^2+2m_S^2)
      +\lambda_{HS}v c_\gamma( c^2_\gamma-2s^2_\gamma)
      +2a_Ss_\gamma c^2_\gamma \label{eq:hSS}  \,,\\
    g_{hhSS} &= \frac{3c_\gamma^2s^2_\gamma}{v^2} (m_h^2 c^2_\gamma+m_S^2s^2_\gamma) +\lambda_{HS}(1-6 c^2_\gamma s^2_\gamma)
    +6\lambda_S c^2_\gamma s^2_\gamma \,, \\
    g_{VVS} &= -\frac{2m_V^2}{v}s_\gamma  \,,\\
    g_{VVSS} &= \frac{2m_V^2}{v^2}s^2_\gamma \,,
\end{align}
with $V=W,Z$, $s_\gamma\equiv \sin\gamma$ and $c_\gamma\equiv \cos\gamma$. 
For the phenomenological analysis, we neglect the singlet self-interactions governed by $a_S$ and $\lambda_S$.

\subsection{Indirect probes}
The model leaves two indirect imprints on the Higgs sector: tree-level coupling shifts induced by the singlet--Higgs mixing, and loop-level shifts induced by the $\lambda_{HS}$ coupling, which survive even at vanishing mixing.
We discuss them in turn.
At tree level, all single-Higgs couplings rescale universally; in particular $\kappa_V=\cos\gamma$ for $V=W,Z$.
For processes at energies below the heavy-scalar mass, the effect of virtual singlet exchange can be described by a low-energy effective theory.
At leading order in the heavy-mass expansion, the relevant operators are
\begin{align}
 \Delta\mathcal L_{\rm EFT}
 &=\frac{c_H}{2\Lambda^2}\big(\partial_\mu|H|^2\big)^2
 -\frac{c_6\lambda_H^{\rm SM}}{\Lambda^2}|H|^6\,,
 &\lambda_H^{\rm SM}&\equiv\frac{m_h^2}{2v^2}\,.
 \label{eq:singlet-EFT-operators}
\end{align}
The normalization $\lambda_H^{\rm SM}$ is fixed by the light-Higgs mass and the electroweak VEV. 
It differs from the UV quartic $\lambda_H$ in \cref{eq:lagdefinition}, which receives a threshold correction when the singlet is integrated out. 
For the benchmark $a_S=\lambda_S=0$, tree-level matching gives
\begin{align}
 \frac{c_H}{\Lambda^2}
 &=\frac{a_{HS}^2}{M_v^4}\,,
 &\frac{c_6\lambda_H^{\rm SM}}{\Lambda^2}
 &=\frac{\lambda_{HS}a_{HS}^2}{2M_v^4}\,,
 &M_v^2&\equiv m_{S,0}^2+\frac12\lambda_{HS}v^2\,.
 \label{eq:singlet-EFT-tree-match}
\end{align}
At small mixing and to leading order in $v^2/m_S^2$, these imply $c_Hv^2/\Lambda^2\simeq\sin^2\gamma$ and $c_6\lambda_H^{\rm SM}/\Lambda^2\simeq \lambda_{HS}\sin^2\gamma/(2v^2)$.
After canonical normalization of the radial Higgs field, keeping $v$ and $m_h$ fixed, the physical Higgs coupling shifts read
\begin{align}
 \Delta\kappa_V
 &=-\frac{c_Hv^2}{2\Lambda^2}
 \simeq-\frac12\sin^2\gamma\,,\nonumber\\
 \Delta\kappa_{V2}
 &=-\frac{2c_Hv^2}{\Lambda^2}
 \simeq-2\sin^2\gamma\,,\nonumber\\
 \Delta\kappa_3
 &=-\frac{3c_Hv^2}{2\Lambda^2}
   +\frac{c_6v^2}{\Lambda^2}
 \simeq\left(-\frac32+\frac{\lambda_{HS}}{2\lambda_H^{\rm SM}}\right)
 \sin^2\gamma\,,\nonumber\\
 \Delta\kappa_4
 &=-\frac{25c_Hv^2}{3\Lambda^2}
   +\frac{6c_6v^2}{\Lambda^2}
 \simeq\left(-\frac{25}{3}+\frac{3\lambda_{HS}}{\lambda_H^{\rm SM}}\right)
 \sin^2\gamma\,.
 \label{eq:kappav}
\end{align}
Here $\Delta\kappa_V$, $\Delta\kappa_{V2}$, $\Delta\kappa_3$ and $\Delta\kappa_4$ denote the relative deviations, $\Delta\kappa\equiv g/g^{\rm SM}-1$, of the $hVV$, $hhVV$, $h^3$ and $h^4$ couplings respectively, following the low-energy coupling normalization of \cite{Fraser:2026ngs}.

Even at vanishing mixing, i.e.~for an exact $\mathbb{Z}_2$ symmetry, where all tree-level shifts above switch off and the scalar is invisible to single-production searches, the $\lambda_{HS}$ coupling still modifies the Higgs couplings at one loop.
The leading matching contributions in this limit are
\begin{align}
 \left.\frac{c_H}{\Lambda^2}\right|_{\rm loop}
 &=\frac{\lambda_{HS}^2}{192\pi^2m_S^2}\,,
 &\left.\frac{c_6\lambda_H^{\rm SM}}{\Lambda^2}\right|_{\rm loop}
 &=\frac{\lambda_{HS}^3}{192\pi^2m_S^2}\,,
 \label{eq:singlet-EFT-portal-loop}
\end{align}
where, at this order in the heavy-mass expansion, the mass in the denominators can be identified with the physical mass $m_S$~\cite{Haisch:2020ahr}.
Inserted into \cref{eq:kappav}, these coefficients capture both one-loop effects of the $\lambda_{HS}$ coupling: the $(\partial_\mu|H|^2)^2$ term acts through the wavefunction renormalization of $h$ and therefore enters every Higgs coupling, while the $|H|^6$ term shifts the self-couplings directly.
The wavefunction piece alone fixes the gauge couplings, 
\begin{equation}
\left.\Delta\kappa_V\right|_{\rm loop}=-\frac{\lambda_{HS}^2v^2}{384\pi^2m_S^2},
\end{equation}
whereas the trilinear coupling receives both contributions,
\begin{align}
 \left.\Delta\kappa_3\right|_{\rm loop}
 &=-\frac{\lambda_{HS}^2v^2}{128\pi^2m_S^2}
 +\frac{\lambda_{HS}^3v^4}{96\pi^2m_h^2m_S^2}\,.
 \label{eq:singlet-kappa3-portal-loop}
\end{align}
This expression retains the leading terms in $\lambda_{HS}$  that survive in the limit $\sin\gamma~\to~0$; the two terms enter with opposite signs for $\lambda_{HS}>0$, and the $|H|^6$ contribution dominates for $\lambda_{HS}\gtrsim0.2$.
At nonzero mixing, additional mixing-dependent one-loop matching
contributions arise, which we do not include.

The gauge couplings of the Higgs are currently measured at the $\sim10\%$ level by the LHC~\cite{ATLAS:2026pdi,CMS:2026zxu}, corresponding in the tree-level mixing interpretation to a $\sin^2\gamma$ of similar magnitude. 
A 10 TeV muon collider is expected to be sensitive to mixings of $\sin^2\gamma\sim10^{-3}$ through the $hWW$ channel \cite{Han:2020pif,Forslund:2023reu}.
The potential of a wakefield collider for Higgs precision studies is explored in~\cite{Fraser:2026ngs}; their mapping to the singlet parameter space is discussed in Section~\ref{sec:conclusions}.

\subsection{Production processes}
The second avenue is the direct discovery of these scalars in a collider environment, which is the focus of this work.  
In the scenario that the stability of $S$ is unprotected by discrete symmetries ($\gamma\neq 0$), it may be produced singly in collider events and decay back into \ac{SM} particles. 
Since these processes are facilitated through the mixing with the Higgs, the production cross section is equivalent to that of a Higgs-like particle with mass $m_S$, weighted by $\sin^2 \gamma$. Like the Higgs, single-$S$ production at high energy colliders is therefore expected to be dominated by \ac{VBF} events.
At a high energy wakefield collider, the luminosity is divided over the $e^+e^-$, $e^-e^-$, $e^\pm\gamma$ and $\gamma\gamma$ initial states. 
The most relevant \ac{VBF} processes for single production are therefore $e^+e^-\to \nu_e\bar\nu_e S$ (\ac{WBF}), $e^-e^-\to e^-e^- S$ (\ac{ZBF}),
$e^-\gamma\to \nu_e W^- S$ ($\gamma e$-induced \ac{WBF}) and $\gamma\gamma\to W^-W^+ S$ ($\gamma\gamma$-induced \ac{WBF}).
In addition one needs to consider next-to-leading order processes such as $\gamma\gamma\to S$ (photon fusion).
The inclusive cross section for all \ac{VBF} processes rises with the center-of-mass energy, thus placing more emphasis on the high energy part of the luminosity spectrum (see \cref{sec:singletsignalprod} for details).
This is a qualitative difference with heavy charged particles and vector resonances, which were previously studied in the context of wakefield colliders \cite{Chigusa:2025azs,Cipressi:2026wtb}.

When single production of $S$ is allowed, it will generically dominate the observable new physics signal, but in the case where the mixing $\sin \gamma$ is extremely small---such as that where the singlet is protected by an approximate (or exact) $\mathbb{Z}_2$ symmetry---or the quartic $\lambda_{HS}$ is extremely large, the production of $SS$ pairs may become important.
The relevant production processes are the same as for single production, upon replacing $S$ by $SS$.
We calculate both the single and pair production cross sections with the MadGraph model from~\cite{Buttazzo:2018qqp}.

\subsection{Decay rates}
$S$ dominantly decays to the weak gauge bosons, and in the heavy scalar limit the Goldstone equivalence theorem and $O(4)$ custodial symmetry would imply $\mathrm{Br}(S\to hh)\simeq \mathrm{Br}(S\to ZZ)\simeq \mathrm{Br}(S\to W^+W^-)/2\sim 1/4$. Among the leading diboson modes, the resonant $hh\to4b$ channel was found to provide the strongest sensitivity at CLIC and a muon collider~\cite{Buttazzo:2018qqp}, motivating the search strategy developed in \cref{sec:singprod}.

Finite masses, singlet--Higgs mixing, and scalar-potential interactions generate departures from this heavy-mass relation.
Defining $x_i=m_i^2/m_S^2$, the trilinear coupling can be parametrized by 
\begin{align}
 \kappa_h
 &\equiv -\frac{v g_{hhS}}{m_S^2\sin\gamma}\nonumber\\
 &=\cos^2\gamma\left[
 1+2x_h
 -\frac{\lambda_{HS}v^2}{m_S^2}(2-\tan^2\gamma)
 -\frac{2a_Sv}{m_S^2}\tan\gamma
 \right].
 \label{eq:kappa-h-general}
\end{align}
Writing $\kappa_h=1+\delta_h$ and expanding in the small quantities $x_h$, $\sin^2\gamma$, $\lambda_{HS}v^2/m_S^2$, and $(a_Sv/m_S^2)\sin\gamma$ gives
\begin{align}
 \delta_h={}&2x_h-\sin^2\gamma
 -2\frac{\lambda_{HS}v^2}{m_S^2}
 -2\frac{a_Sv}{m_S^2}\sin\gamma
 +\mathcal O(\epsilon^2)\,,
 \label{eq:delta-h-expansion}
\end{align}
where $\epsilon$ collectively denotes these small quantities.
In the limit that $\lambda_{HS}=a_S=0$, the leading two-body widths are
\begin{align}
 \Gamma(S\to hh)&=\frac{m_S^3\sin^2\gamma}{32\pi v^2}
       \kappa_h^2\sqrt{1-4x_h}\,,\nonumber\\
  \Gamma(S\to ZZ)&=\frac{m_S^3\sin^2\gamma}{32\pi v^2}
       (1-4x_Z+12x_Z^2)\sqrt{1-4x_Z}\,,\nonumber\\
  \Gamma(S\to W^+W^-)&=\frac{m_S^3\sin^2\gamma}{16\pi v^2}
       (1-4x_W+12x_W^2)\sqrt{1-4x_W}\,,
    \label{eq:singlet-heavy-widths}
\end{align}
which, for example, lead to the two-Higgs branching ratio
\begin{align}
 \mathrm{Br}(S\to hh)&\simeq
 \frac{\kappa_h^2\sqrt{1-4x_h}}
 {\kappa_h^2\sqrt{1-4x_h}+2F(x_W)+F(x_Z)}\,,
 \label{eq:BRShh-analytic} \\
 F(x)&=(1-4x+12x^2)\sqrt{1-4x}\,.
\end{align}
In the same small-parameter expansion, \cref{eq:BRShh-analytic} gives
\begin{align}
 \mathrm{Br}(S\to hh)\simeq\frac14\left[1+\frac32\delta_h-\frac32x_h
 +3x_W+\frac32x_Z+\mathcal O(\delta_h^2,x_i^2,\delta_h x_i)\right].
 \label{eq:BRShh-expansion}
\end{align}
This expression shows that, for the small mixings relevant to our projected reach, the $1/4$ approximation is sound.
At $m_S=1\,\mathrm{TeV}$ the relative difference from the full result is 1.9\%, and for $m_S\gtrsim2\,\mathrm{TeV}$ it is less than $1\%$.\footnote{In evaluating these branching ratios, we also include the decay $S\to t\bar t$.
This fermionic width is Yukawa suppressed: $\Gamma_{t\bar t}\propto m_Sm_t^2/v^2$, whereas the longitudinal-vector and di-Higgs widths scale as $m_S^3/v^2$, giving a relative suppression of order $m_t^2/m_S^2$.
For small but nonzero mixing, neglecting this contribution shifts $\mathrm{Br}(S\to hh)$ from $24.5\%$ to $26.4\%$ at $m_S=\SI{1}{\TeV}$ and from $24.9\%$ to $25.1\%$ at $m_S=\SI{4}{\TeV}$.
}
Even for the larger illustrative value $\sin\gamma=0.1$, the deviation relative to $1/4$ ranges only from $3.4\%$ at $1\,\mathrm{TeV}$ to $1.6\%$ at $8\,\mathrm{TeV}$.
We therefore use $\mathrm{Br}(S\to hh)=1/4$ in the analysis.
This validation is benchmark dependent since sizable $\lambda_{HS}$ or $a_S$ can modify the $hhS$ coupling, but is sufficient for the single-production case. 

  The quality of the $\mathbb{Z}_2$ symmetry in \eqref{eq:lagdefinition} will determine the lifetime of $S$, and its decay rate is therefore suppressed by $\sin^2\gamma$ 
\begin{equation}
    c \tau_S \approx \, 3\text{ mm }\times \left(\frac{m_S}{\text{TeV}}\right)^{-3} \,\times\, \left(\frac{\sin\gamma}{10^{-8}}\right)^{-2}.
\end{equation} 
For a TeV-scale scalar singlet, mixings below $ \sin \gamma \sim 10^{-11}$ will result in a particle that is long-lived with respect to the detector, while a mixing angle above $\sim 10^{-8}$ will ensure a prompt decay.
Mixing angles in the intermediate regime will result in decays with displaced vertices.
In the latter two cases, the new physics signature should be a distinctive one.

Finally, in the regime where pair production is dominant, the branching ratios may be substantially deformed as the quartic becomes large, $\lambda_{HS} \gtrsim \mathcal{O}(1)$.
As shown in Eq.~\eqref{eq:kappa-h-general}, the $hhS$ coupling approximately vanishes as $\lambda_{HS} = m_S^2/2v^2$, while the vector couplings remain unchanged, leading to a narrower width and $\sim100\%$ branching ratio into $W^\pm$ and $Z$ bosons.
For $m_S \sim\SI{1}{TeV}$ this occurs at $\lambda_{HS} \sim 3\pi$.
It should be noted that such large values of $\lambda_{HS}$ may threaten the stability of the scalar sector potential unless extra parameters (such as the quartic $\lambda_S$) are turned on, as $m_{S,0}^2 \approx m_S^2 - \lambda_{HS} v^2/2$ becomes negative for large $\lambda_{HS}$.
In this work, we restrict our discussion in this regime to the purely phenomenological study of $SS$ pair production, detailed in \cref{sec:singletpairprod}.

\section{Single production\label{sec:singprod}}

\subsection{Analysis strategy\label{sec:singleSstrat}}
\begin{figure}
    \centering
    \includegraphics[width=0.6\linewidth]{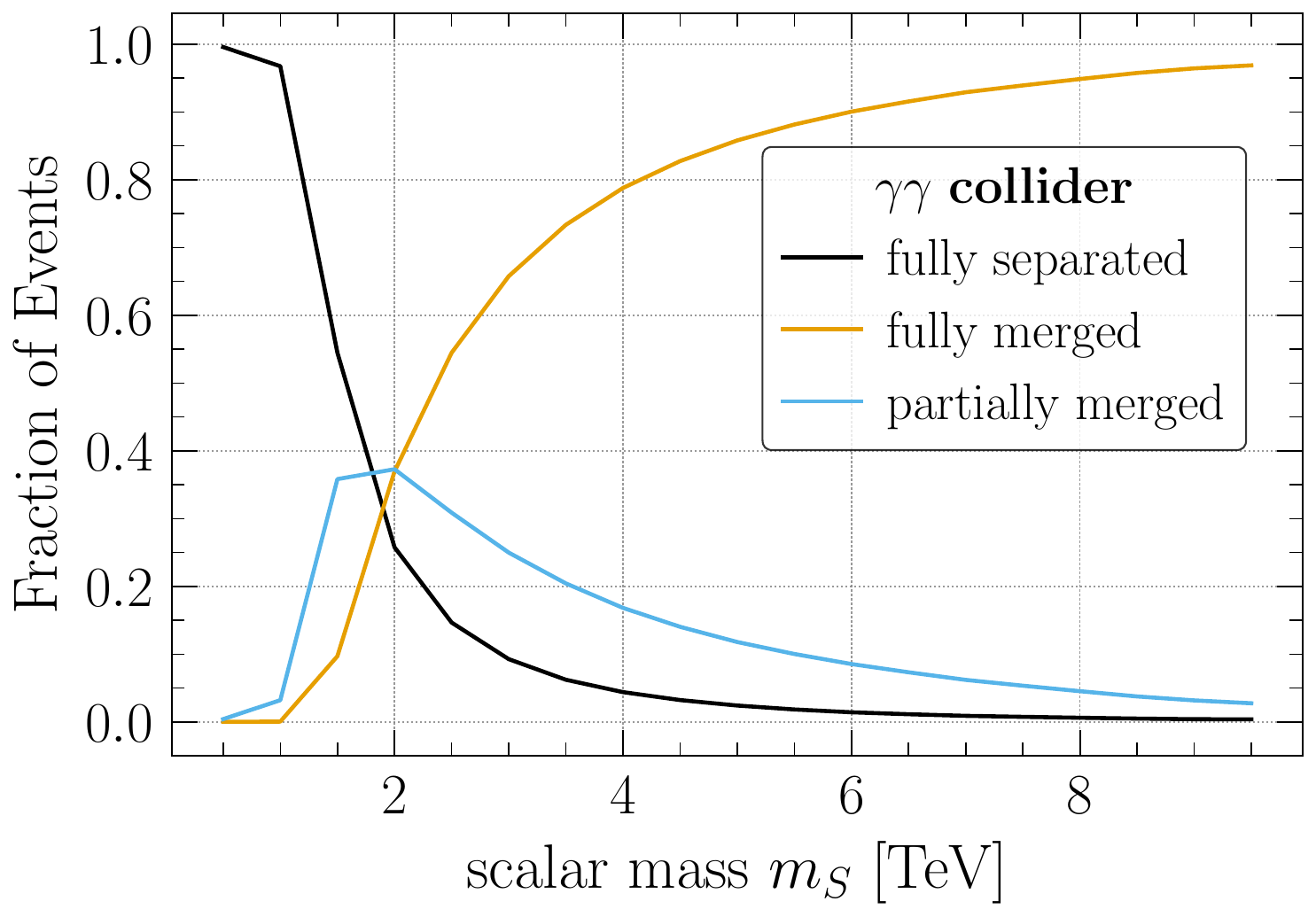}
    \caption{Fractions of events that fall into the fully separated, fully merged and partially merged categories at a $\gamma\gamma$ collider. 
    }
    \label{fig:categories}
\end{figure}

As explained in \cref{sec:singletmodel}, we focus on the $S\to hh\to 4b$ decay mode. 
At a \SI{10}{\TeV} collider, each $h$ can be boosted enough for its $b\bar b$ pair to merge into a single jet.
Merging changes which backgrounds dominate, so we sort the events into three mutually exclusive categories based on the angular separation of the two $b$ quarks from each Higgs decay.
We quantify this separation using the standard hadron-collider variable $\Delta R_{bb}\equiv\sqrt{\Delta\phi^2+\Delta\eta^2}$, where $\Delta\phi$ and $\Delta\eta$ are the differences in azimuthal angle and pseudorapidity of the two $b$ quarks, respectively.
The categories are:
\begin{itemize}
    \item Fully separated: At truth-level, all $b$-quarks are separated by $\Delta R_{bb}>0.4$ from one another. For this case we assume that they all can be reconstructed as separate jets.
    \item Fully merged: At truth-level, all $b$-quarks have a separation of $\Delta R_{bb}<0.4$ from the nearest other $b$-quark. Here we assume that those pairs are reconstructed as a single ``double-$b$'' jet. We require that those double-b jets are always well separated from each other.
    \item Partially merged: One pair of $b$-quarks is merged ($\Delta R_{bb}<0.4$) and the other pair is not ($\Delta R_{bb}>0.4$). We also require that the double-b jet is always well separated from the two regular $b$-jets.
\end{itemize}
In \cref{fig:categories} we show how the different categories are populated as a function of $m_S$ for the $\gamma\gamma$ collider.
Very little difference in the event fractions is found between the different colliders we consider.

For the fully separated category, the main backgrounds are $hh$, $Zh$ and $ZZ$ production, where we can effectively reduce the latter two with a cut on the invariant mass of the $b\bar b$ pair. 
For the fully merged category, the $hh$, $Zh$ and $ZZ$ channels also contribute in the kinematic regime where the $h$ and/or $Z$ are highly boosted. 
In addition, $b\bar b$ and $c\bar c$ production must be considered: The cross sections for these processes are very high and the fragmentation functions can occasionally generate a $b$/$c$-jet with a high enough invariant mass to be consistent with a merged $h\to b\bar b$ decay.  
To further reduce this background, we assume a dedicated tagging algorithm that can separate a $b$/$c$-jet from a double-$b$ jet to a sufficient degree (see \cref{sec:singletanalysiscuts} and \cref{app:tagger} for details).
For the partially merged category, we apply this tagger's efficiencies to the merged $b\bar b$ pair and the $m_{b\bar b}$ cut on the separated $b\bar b$ pair.
For all categories, we impose cuts on the invariant mass ($m_{4b}$) and total transverse energy ($H_T^{4b}$) of the four-$b$ system, see \cref{sec:singletanalysiscuts}.

\subsection{Simulation framework\label{sec:detector}}

We compare five collider configurations: two $e^+e^-$ colliders and two $e^-e^-$ colliders, each with flat and round beams, and a $\gamma\gamma$ collider.
The beam parameters result from a preliminary optimization performed by the LBNL BELLA group \cite{Bulanov:2026nqa,Formenti:2026jun,Schroeder:2022xdu}.
We use the integrated geometric luminosity
\begin{align}
    \Lumgeom &\equiv \frac{N_e^2 \nu_\text{rep} T}{4\pi \sigma_x \sigma_y}\,,
\end{align}
as a common basis for comparing collider designs.
Here $N_e$ is the number of electrons per bunch, $\nu_{\text{rep}}$ the repetition rate, $T$ the run time of the facility and $\sigma_{x,y}$ the beam sizes in the transverse directions.
The quantity $\Lumgeom$ is the luminosity the accelerator delivers to the \ac{IP} in the absence of beam--beam interactions, and is most analogous to the luminosity quoted for muon or hadron colliders.

The beam--beam interactions however strongly disrupt the crossing beams, causing them to radiate hard photons, which in turn produce $e^+e^-$ pairs via the Breit-Wheeler process.
Calculations of particle production rates therefore require the full, energy-dependent luminosity spectra rather than just $\Lumgeom$.
These spectra are analogous to parton distribution functions, but with two important differences: all particles are real rather than virtual, and the spectra cannot be factorized into a product of single-beam distributions.\footnote{The contribution from electroweak parton distribution functions is typically a negligible correction compared to the beam--beam interactions, see \cite{Chigusa:2023rrz,Cipressi:2026wtb}.}
Their shape depends on collective beam parameters such as $\sigma_{x,y}$ and $N_e$, directly connecting the physics reach to the accelerator design.
The interaction region simulations were carried out by A.~Formenti et al.~in the LBNL Advanced Modeling Program \cite{Formenti:2026jun}.
\texttt{WarpX} \cite{warpx} was used for the $e^+e^-$ and $e^-e^-$ colliders.
The design of the interaction region for the $\gamma\gamma$ collider was optimized in \cite{Bulanov:2026nqa}, with the Compton conversion and photon collisions handled by \texttt{CAIN} \cite{Chen:1994jt}, using the same incoming $e^-e^-$ round-beam parameters.
These simulations produce the luminosity spectra in \cite{Formenti:2026jun}. 
Integrated over $M\in[\SI{0.5}{\TeV},\SI{10}{\TeV}]$, these spectra amount to approximately $2.5\,\Lumgeom$ ($15\,\Lumgeom$) in $e^+e^-$ collisions for flat
(round) beams, as the oppositely charged bunches pinch each other during the crossing, and $\simeq0.7\,\Lumgeom$ for the anti-pinching $e^-e^-$
configurations; see Table~1 of \cite{Chigusa:2025azs} for all initial states. 

All signal and background event rates are modeled with \texttt{MadGraph5\_aMC@NLO}~\cite{Alwall:2014hca}.
Following \cite{Chigusa:2025azs}, we do so on a regular grid in the collision center-of-mass energy $M$ with \SI{50}{\GeV} intervals and reweight the events using the luminosity spectra for the initial state in question.
All $Z$ and $h$ bosons are decayed to $b$ quarks, which we require to have a lab-frame pseudorapidity $|\eta_b|<3$.
When applying this cut we account for the energy imbalance between the incoming particles by folding in the probability distribution of the center-of-mass boost relative to the lab frame, as extracted from the \texttt{WarpX} and \texttt{CAIN} simulation data. 

Rather than running a full parton shower and detector simulation, we instead apply a Gaussian smearing to the truth-level $b$-quark momenta in order to capture the effects of jet energy resolution.
For the jet energy resolution we take $\delta E/E= 10\%$, and treat the angular components of the jet momenta as having negligible uncertainty. 
At the $Z$-pole, this corresponds to an uncertainty on the invariant mass of the $b$-quarks ($m_{bb}$) of $\delta m_{bb}/m_{bb}\approx 7.5\%$.
This resolution is the key factor in how well $Z$ bosons can be distinguished from Higgs bosons, and therefore directly affects the background rates.
For comparison, ATLAS achieved an $m_{bb}$ resolution of about $10\%$ for $h\to b\bar b$ decays in Run~2~\cite{ATLAS:2017cen}, and CLIC projects a resolution of about $7\%$ for hadronic $Z$ decays at its \SI{3}{\TeV} stage, including the expected beam-induced $\gamma\gamma\to$ hadrons background~\cite{CLICdp:2018vnx}.
This background is expected to be more severe at a \SI{10}{\TeV} wakefield collider, particularly with round beams. 
By convolving the $\gamma\gamma\to$ hadrons rate \cite{Barklow:1443518} with the luminosity spectra, we find a collision environment containing $\mathcal{O}(1-10)$ times more soft tracks than at CLIC, depending on the collider in question. 
Our choice of an $m_{bb}$ resolution somewhat better than at the LHC, and comparable to the CLIC projection~\cite{CLICdp:2018vnx}, thus assumes that advances in analysis techniques and detector technology will compensate for these larger beam-induced backgrounds.
As we will see in \cref{sec:backgrounds}, our conclusions do not strongly depend on this assumption.

For $b$-jets that are at least $\Delta R_{bb}>0.4$ separated from all other $b$-jets in the event, we assume an aggressive $b$-tagger operating at 90\% efficiency.
At such a working point, ATLAS has achieved a mistag rate of roughly 50\% for $c$-jets and around 10\% for light flavors \cite{ATLAS:2022qxm}.
A comparable performance is currently anticipated for CLIC \cite{CLICdp:2018vnx}, with the possibility of further improvement through modern machine learning techniques.
Adopting a $50\%$ charm mistag rate implies that approximately $25\%$ of all $Z\to c\bar c$ decays would be incorrectly identified as $Z\to b\bar b$ decays.
We account for this effect by applying the appropriate correction factor to these backgrounds.
The contribution from mistagged light flavor jets is found to be negligible.

If two $b$-jets are close together ($\Delta R_{bb}<0.4$), we assume that they are reconstructed as a single jet. 
In this case, we assume that a specialized tagging algorithm will be able to separate jets with two $b$-quarks from jets with a single $b$-quark.
We assume such an algorithm to have a signal efficiency of 80\% and a single-$b$ to double-$b$ mistag rate of 2\%.
We validate this assumption in \cref{app:tagger}.

\subsection{Signal processes\label{sec:singletsignalprod}}

\begin{figure}[t]
    \centering
    \includegraphics[width=\linewidth]{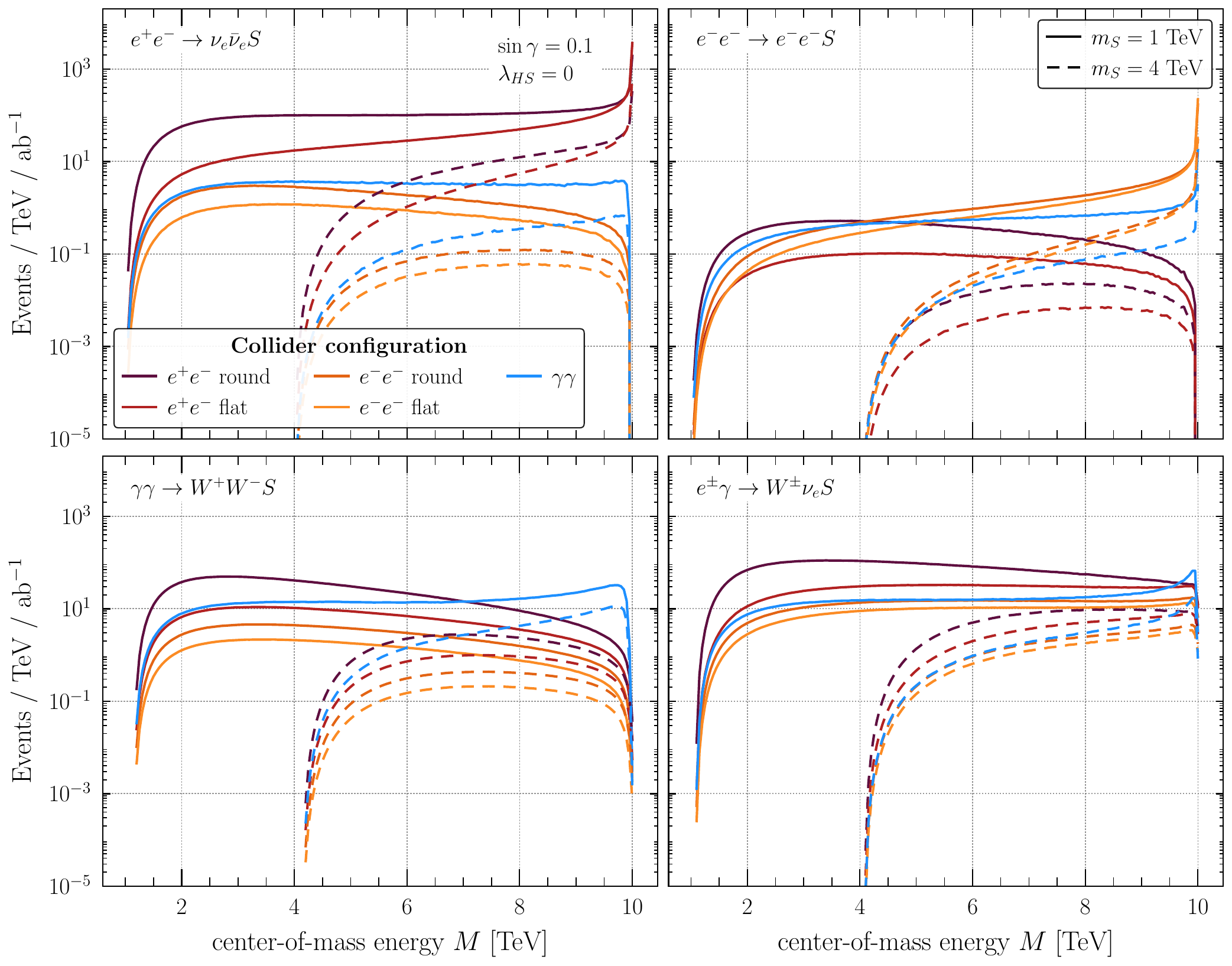}
    \caption{
        Numbers of inclusive single-$S$ signal events as functions of the collision center-of-mass energy $M$ for the benchmark scalar masses $m_S=\SI{1}{\TeV}$ (solid) and $\SI{4}{\TeV}$ (dashed). 
        The four panels correspond to the four processes we consider in our analysis, $e^+e^- \to \nu_e \bar{\nu}_e S$, $e^-e^- \to e^-e^- S$, $\gamma\gamma\to W^+W^- S$ and $e^\pm\gamma\to W^\pm \nu_e S$; the different colors correspond to the collider options.
        }
    \label{fig:signal_events_differential}
\end{figure}

In the multi-TeV regime, single-$S$ production is dominated by vector-boson-fusion topologies: in particular, the charged-current channel $e^+e^-\to\nu_e\bar{\nu}_eS$ induced by \ac{WBF} is expected to be substantially larger than the neutral-current channel $e^\pm e^-\to e^\pm e^-S$ induced by \ac{ZBF}, reflecting the larger electroweak coupling of the $W$ to leptons.
A distinctive feature of a \SI{10}{\TeV} wakefield collider is its large $e^\pm\gamma$ and $\gamma\gamma$ luminosity, which opens up the additional $W$-fusion processes $e^-\gamma\to\nu_e W^- S$ and $\gamma\gamma\to W^+W^-S$.
These channels provide the leading contributions to single-$S$ production at the $e^-e^-$ and $\gamma\gamma$ colliders.
Summed over all \ac{VBF} channels, the single-$S$ cross section grows logarithmically with the center-of-mass energy. 

The large $\gamma\gamma$ luminosity also allows photon fusion, $\gamma\gamma\to S$, whose cross section in the narrow-width limit and for unpolarized photons is
\begin{equation}
  \sigma(\gamma\gamma\to S)=\frac{4\pi^2}{m_S^2}\,
  \Gamma(S\to\gamma\gamma)\,\delta(M-m_S)\,,
\end{equation}
where $M$ is the $\gamma\gamma$ invariant mass. 
The partial width $\Gamma(S\to\gamma\gamma)$ is $\sin^2\gamma$ times the \ac{SM} expression for $\Gamma(h\to\gamma\gamma)$ (see e.g.~\cite{Djouadi:2005gi}), with $m_h$ replaced by $m_S$.
Finally, the three-body process $\gamma\gamma\to t\bar tS$ has an amplitude proportional to the top Yukawa coupling times the singlet--Higgs mixing.
Despite the large top Yukawa coupling, its cross section remains much smaller than those of the \ac{VBF} and photon-fusion channels, so we neglect it.

\begin{figure}
    \centering
    \includegraphics[width=0.95\linewidth]{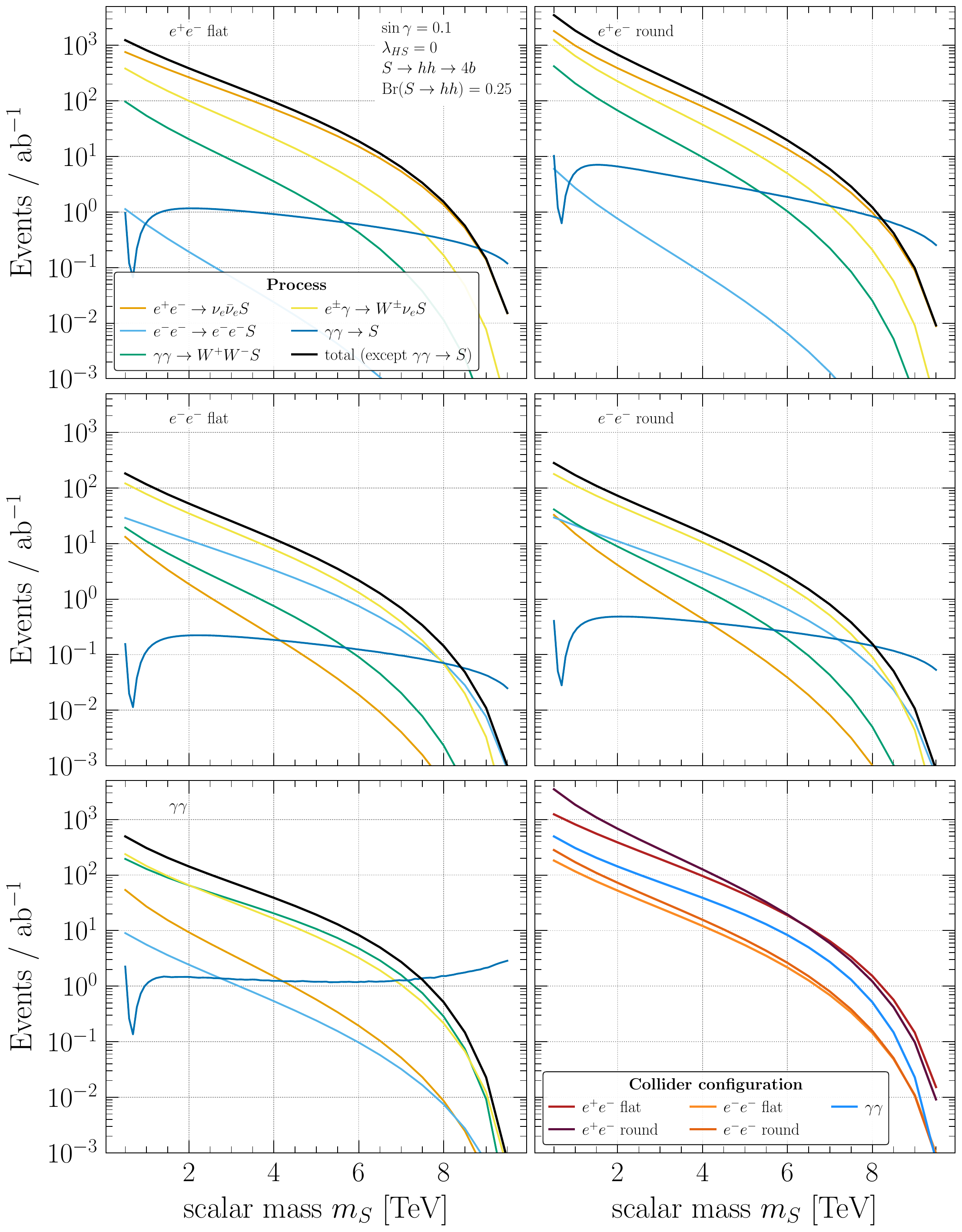}
    \caption{Total numbers of inclusive single-$S$ signal events as functions of the scalar mass $m_S$.
    Five panels correspond to the five collider configurations considered.
    Each panel shows the contributions to the total \ac{VBF} signal rate (black) of the four processes we consider in our analysis  $e^+e^- \to \nu_e \bar{\nu}_e S$ (orange), $e^-e^- \to e^-e^- S$ (light blue), $\gamma\gamma\to W^+W^- S$ (green) and $e^\pm\gamma\to W^\pm \nu_e S$ (yellow), as well as the analytic rate for the process $\gamma \gamma \to S$ (dark blue).
    The bottom right panel compares the total signal rate (not including $\gamma\gamma\to S$) for all five collider configurations.
   The branching ratios for the full decay chain are included, $\mathrm{Br}(S\to hh)\,\mathrm{Br}(h\to b\bar b)^2\simeq 0.08$. }
    \label{fig:signal_events_inclusive_mass}
\end{figure}

The dependence of the event rates on the collision energy is shown in \cref{fig:signal_events_differential}. The  different line types correspond to different singlet masses;
the different panels represent the four \ac{VBF} production processes. The differential distribution of the photon fusion process is a $\delta$-function and therefore not plotted. 

Despite the different cross section scaling of the \ac{VBF} processes compared to the production of heavy charged particles and vector resonances, some of the same features are observed: the \ac{VBF} processes $e^+e^-\to \nu_e\bar{\nu}_e S$ and $e^-e^-\to e^-e^-S$ feature a sharp peak at the endpoint of the distribution for $e^+e^-$ and $e^-e^-$ collider options, respectively.
This directly mirrors the shape of the corresponding luminosity distributions.
Similarly, there is a sharp peak for $e^-e^-\to e^-e^-S$ at a photon collider as only a fraction of the primary beam electrons are converted to photons via Compton-backscattering.
This also leads to the less pronounced peaks in the photon-induced \ac{WBF} processes $\gamma\gamma \to W^-W^+ S$ and $e^-\gamma \to \nu_e W^- S$ at $M=\SI{10}{\TeV}$ for a $\gamma\gamma$ collider.
Specifically, $e^-\gamma\to \nu_e W^- S$ benefits from the surviving electrons, leading to a more pronounced peak.
While it is not as prominent as for resonant production (see \cite{Chigusa:2025azs,Cipressi:2026wtb}), in addition to the production at peak energies a large fraction of scalars is produced at lower $M$ for all four \ac{VBF} processes at all five collider options.
This is most prominent for round $e^+e^-$ beams where strong beam-beam interactions lead to large luminosities at low collision energies.
The contributions from the low energy luminosity tails are particularly relevant for $e^-e^-$ and $\gamma\gamma$ collider options, where they can partially compensate the lower peak luminosities of the main production modes.

\Cref{fig:signal_events_inclusive_mass} shows the dependence on the singlet mass; the event yields fall rapidly with increasing $m_S$.
In addition to the four \ac{WBF} and \ac{ZBF} processes discussed above the event yield for photon fusion is also shown.
It only becomes the dominant production mode for $m_S\gtrsim\SI{8.5}{\TeV}$, except of the $\gamma\gamma$ collider, where it can dominate for slightly lower $m_S$.
In the high $m_S$ region, photon fusion contributes at most $\mathcal{O}(10)$
events per \SI{10}{\per\atto\barn} at the $\gamma\gamma$ collider and a few
events or fewer at the other collider options, and it remains a subdominant
correction to the total rate except near the kinematic limit, where the total
expected yield is itself at the few-event level.
Therefore, it is not included in the total event count shown as black lines for each collider, and summarized in the lower right panel.
As expected the main production channel in both $e^+e^-$ collider options is \ac{WBF} with a sub-leading contribution from $e^-\gamma$-induced \ac{WBF}.
Round beams lead to a slightly higher event yield for scalar masses $m_S\lesssim \SI{7}{\TeV}$ as they benefit from the larger low energy luminosity tails. 
Indeed, the round and flat $e^+e^-$ spectra contain equal integrated luminosity in the top 20\% of the kinematic range, while the factor of six surplus of the round configuration sits entirely at lower $M$.  However, this difference in low-energy luminosity is partially undermined by the energy-dependence of the \ac{VBF} cross section, which scales logarithmically with the center-of-mass energy and brings the round and flat beam performances much closer together. An illustrative comparison is the signal event yields of charged particle production studied in~\cite{Chigusa:2025azs}, where the same higher luminosity tails result in a substantially larger difference in low mass particle production between round and flat beams, compared to what is shown in~\cref{fig:signal_events_inclusive_mass}.
For $e^-e^-$ beams $e^-\gamma$-induced \ac{WBF} is the main production mode up to $m_S\sim\SI{9}{\TeV}$ where \ac{ZBF} becomes dominant because of the luminosity peak at $M=\SI{10}{\TeV}$.
Finally, for the $\gamma\gamma$ collider the two photon-induced \ac{WBF} processes contribute about equally.
Overall, assuming the same geometric luminosity, $e^+e^-$ round beams produce the most signal events for $m_S\lesssim\SI{7}{\TeV}$ and $e^+e^-$ flat beams for larger scalar masses.
The $\gamma\gamma$ collider always produces more signal events than both $e^-e^-$ options.

\subsection{Background processes\label{sec:backgrounds}}

\begin{figure}[t]
    \centering
    \includegraphics[width=\linewidth]{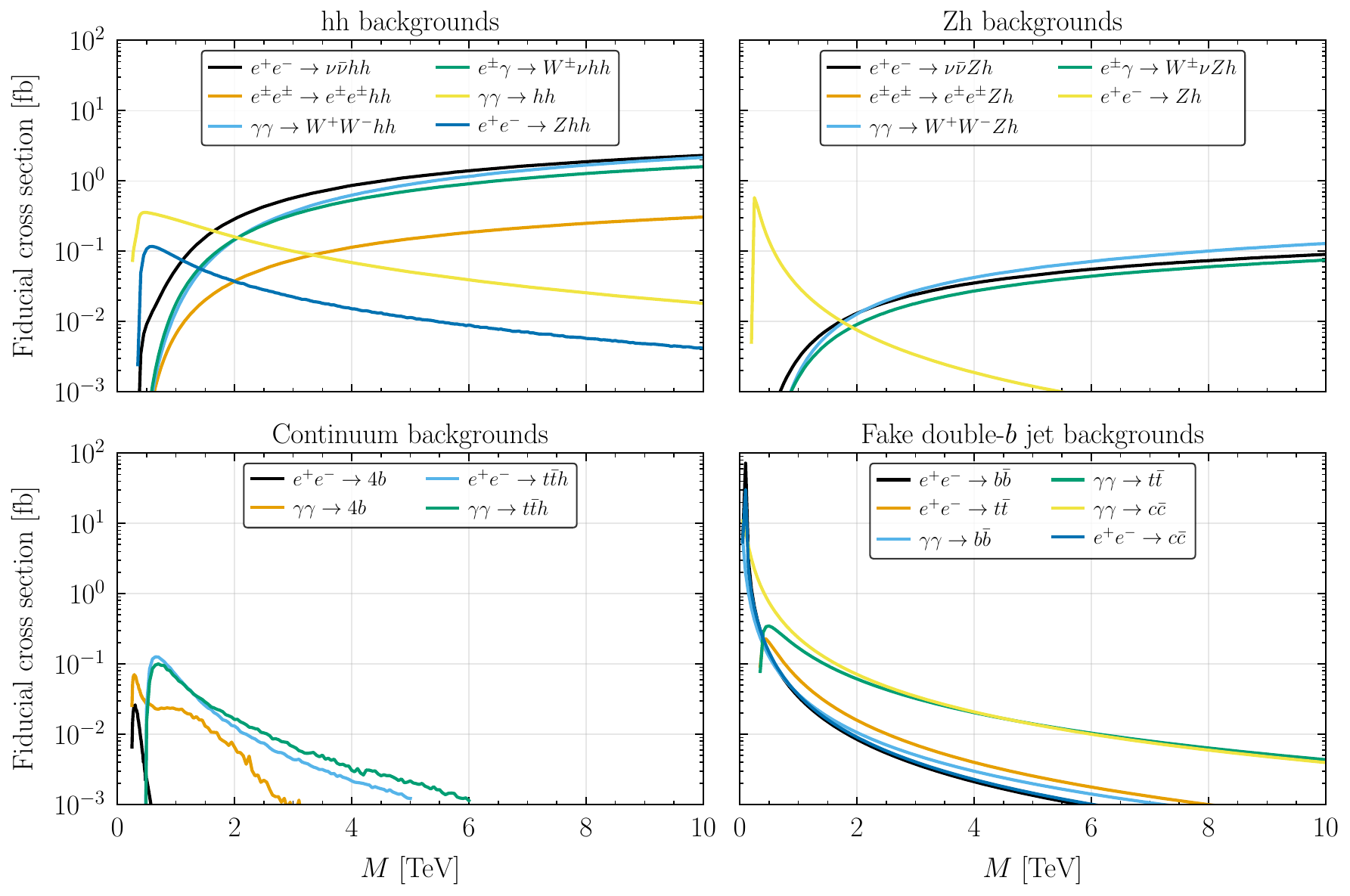}
    \caption{Fiducial cross sections of the most important background processes multiplied by the branching ratios for the $h$ and $Z$ decays to $b\bar b$. 
    For the $Zh$, continuum and fake double-$b$ jet backgrounds, additional cuts and/or scale factors have been applied to estimate the effect of the analysis cuts in \cref{sec:singletanalysiscuts}.
    See text for details.
    }
    \label{fig:backgroundxsec}
\end{figure}

The backgrounds can be classified into four categories:
\begin{enumerate}[label=\emph{\roman*)}]
    \item Final states with genuine $hh$ pairs.
    \item Final states in which one or more $Z\to b\bar b$ or $Z\to c\bar c$ decays are misidentified as $h\to b\bar b$.
    \item Continuum backgrounds, where the invariant mass of two a priori unrelated $b$-jets approximately matches $m_h$. These include the $4q$ continuum and $t\bar t h$ production.
    \item For ``merged'' and ``partially merged'' categories we account for $c\bar c$, $b\bar b$ and $t\bar t$ processes, as individual $c$/$b$-quarks can fake a double-$b$ jet consistent with a boosted $h\to b\bar b$ decay. 
\end{enumerate}
The cross sections for all backgrounds are summarized in \cref{fig:backgroundxsec} for select values of $M$, including the corresponding branching fractions, with $\mathrm{Br}(h\to b\bar{b})\simeq 0.58$ and $\mathrm{Br}(Z\to b\bar{b})\simeq 0.15$ folded in as appropriate.
To make their comparison more instructive, some baseline selection cuts or reweighting have also been applied to some processes, as explained below.

\Ac{VBF} processes supply most of the $hh$ backgrounds through a combination of the $e^+e^-$, $\gamma\gamma$ and $e^-\gamma$ initial states (upper left panel in \cref{fig:backgroundxsec}).
As for the signal processes, they grow at high energy.
All $hh$ processes that involve \ac{VBF} have comparable cross sections around a few fb at 10 TeV, except for \ac{ZBF} ($e^\pm e^\pm \to e^\pm e^\pm h h$).
The latter is considerably smaller due to the relatively small coupling of the $Z$ boson to the charged leptons in the Standard Model.
The relative importance of the individual \ac{VBF} processes depends on the collider under consideration.
We take them all into account in our analysis, weighted by the luminosity spectra for each collider.
The direct di-Higgs channels at one-loop level do not benefit from a logarithmic high-energy enhancement characteristic of vector-boson fusion and instead fall rapidly at higher energies.
The $\gamma\gamma \to hh$ process is nevertheless dominant for $M<\SI{1}{\TeV}$ and is therefore included accordingly.
A priori, the loop-induced $e^+e^- \to hh$ also contributes, but it only has a cross section of the order of $\sim \SI{10}{\atto\barn}$ \cite{Lopez-Villarejo:2008qii,Kriewald:2025bui} and is therefore neglected and not shown in \cref{fig:backgroundxsec}.
Finally, we consider the $e^+e^-\to Zhh$ process.
Its cross section is strictly smaller than that of $\gamma\gamma \to hh$; however, this is offset by the larger $e^+e^-$ luminosity at low $M$, which requires us to include it in the total background count.

Before cuts, the $Zh$ rates are comparable to or somewhat higher than the $hh$ rates, but the $Zh$ contribution can be suppressed by requiring both $b\bar b$ pairs to have invariant masses close to $m_h$ after detector smearing. 
For our final results we place this cut directly on the smeared Monte Carlo events, as described in \cref{sec:singletanalysiscuts}.
To provide a quantitative estimate of the importance of the $Zh$ background, we have rescaled the cross sections with an estimate of the probability $p^{\text{fake}}_Z$ that a $Z\to b\bar b$ decay can fake an $h\to b\bar b$ decay. 
For resolved $b\bar b$ pairs, the Higgs mass window $|m_{bb}-m_h|<\SI{15}{\GeV}$ and a $7.5\%$ Gaussian mass resolution give a $Z\to b\bar b$ acceptance of $p^{\text{fake}}_Z\simeq0.003$ when the truth-level $m_{bb}$ is fixed to $m_Z$. 
Including the finite width of the $Z$, we estimate $p^{\text{fake}}_Z\simeq0.02$. 
The $Zh$ cross sections in the upper right panel in \cref{fig:backgroundxsec} are rescaled using the $p^{\text{fake}}_Z\simeq0.02$ value.
From this estimate, we expect this background to always be negligible compared to the $hh$ background.
We confirm this with a set of full Monte Carlo simulations: The $Zh$ background rates used in our analysis are obtained from  Monte Carlo events with the $Z$-width properly accounted for and the jet energies smeared as described in~\cref{sec:detector}, so that $m_{bb}$ has a $7.5\%$ detector resolution. 
Then the analysis cuts are applied to this smeared sample.
For merged candidates, we conservatively assign $Z$ and Higgs decays the same tagging efficiency.
The $Zh$ background is accounted for in all categories, but in practice it does not affect the sensitivity. Finally, we verified that all $ZZ$ backgrounds are negligible once the $(p^{\text{fake}}_Z)^2$ suppression factor is accounted for.

For the continuum, we consider the $4b$ background from $e^+e^-\to 4b$ and $\gamma\gamma\to 4b$ processes.
The latter's cross section is much larger due to a $t$-channel enhancement.
We required at truth-level that $p_T>\SI{30}{\GeV}$ for all $b$-quarks, that there exist two $bb$ pairings in the event which each satisfy $|m_h-m_{bb}|<\SI{15}{\GeV}$ and that $|\eta_{b_i}-\eta_{b_j}|<6$ for all pairwise combinations of the $b$-quarks.
This final criterion removes the forward part of the $\gamma\gamma \to 4b$ phase space, which is not covered by our assumed detector configuration.
The resulting fiducial cross section is shown in the bottom left panel in \cref{fig:backgroundxsec}.
We find that $e^+e^-\to 4b$ is always negligible and that $\gamma\gamma\to 4b$ only contributes for $m_{4b}<1$ TeV.
It is nevertheless still small compared to the $hh$ backgrounds and is therefore neglected. 
Similarly, we consider $4c$ and $2b+2c$ continuum backgrounds, which have higher electroweak cross sections than the $4b$ background.
However, with a charm mistag rate of $\sim50\%$, these backgrounds are smaller than or comparable to the $4b$ background, and therefore also negligible.
The $e^+e^-\to t\bar t h$ and $\gamma\gamma\to t\bar t h$ processes also contribute in cases where the two $b$-quarks from the $t$ decays accidentally reconstruct $m_h$ to within the resolution of the detector.
In \cref{fig:backgroundxsec} we again selected for this part of the phase space by requiring $|m_h-m_{bb}|<\SI{15}{\GeV}$.
The cross sections for both processes are small compared to those of the $hh$ backgrounds, but there are nevertheless particular corners of phase space where they may be relevant.
For example, for the $e^+e^-$ round collider and $M<\SI{1}{\TeV}$, $e^+e^-\to t\bar t h$ contributes comparably to $\gamma\gamma\to hh$ and exceeds all \ac{VBF} rates.
We therefore include these background processes in our analysis in the next section.

Finally, for the merged analysis category we must consider processes where single $c$-jets or $b$-jets are misclassified as jets from a boosted $h\to b\bar b$ decay.
The most important processes contributing $b$ quarks are $b\bar b$ and $t\bar t$ production with the $e^+e^-$ and $\gamma\gamma$ initial states, which each have cross sections that are orders of magnitude higher than those of the $hh$ backgrounds.
In addition, we must consider $\gamma\gamma\to c\bar c$ and $e^+e^-\to c\bar c$ whose cross sections are roughly a factor of 16 and a factor of $2$ larger, respectively, than their $b\bar b$ counterparts in the $M\to \infty$ limit. 
It is therefore important to reliably distinguish single-$b$ and single-$c$ jets from boosted $h\to b\bar b$ jets. 
With an excellent detector this is possible, but never perfect.
This is because the fragmentation process for a single $b/c$ quark occasionally produces a second $b$-quark or a $c$-quark that cannot be distinguished from a second $b$ quark from $h$ decay.
In \cref{app:tagger}, we estimate that $h\to b\bar b$ decays can be tagged with a probability of 80\%, with a fake rate from $b$ ($c$) jets of about $1/52$ ($1/69$).
Much better $c$-quark suppression is certainly possible;  we chose a working point for the tagger that prioritizes signal efficiency over $c$-quark rejection, which is what maximizes the sensitivity.
For the cross sections in the bottom right panel of \cref{fig:backgroundxsec} we applied this penalty for both $b/c$ jets in the event.
We see that $\gamma\gamma \to t\bar t$ and $\gamma\gamma \to c\bar c$ have the highest effective cross sections. However, after applying the analysis cuts in \cref{sec:singletanalysiscuts} we find that the largest irreducible contributions come from $e^+e^- \to c\bar c$, followed by $e^+e^- \to b\bar b$. 
Both are ultimately negligible compared to the genuine $hh$ backgrounds.

\subsection{Analysis cuts\label{sec:singletanalysiscuts}}

The baseline selection cuts are
\begin{equation}
    p_T> \SI{30}{\GeV}, \quad |\eta|<3 \quad \forall\, b\text{-jets}.
\end{equation}
To capture all signal production modes in \cref{sec:singletsignalprod}, we opt for an \textit{inclusive analysis} and do not impose any further selection on additional particles that may be produced in association with $S$. The core discriminating power between signal and backgrounds comes from the nested resonance structure of the $S\to hh\to4b$ signature.
After smearing the energy of the $b$ quarks, we group them into two Higgs candidates with invariant masses near $m_h$, by choosing the pairing that minimizes $|m_{2b}^{(1)}-m_h|+|m_{2b}^{(2)}-m_h|$. 
We then compute the $\Delta R_{bb}$ for each pair and place the event in the fully separated, fully merged or partially merged category, as introduced in \cref{sec:singleSstrat}. For each category, further analysis cuts are then applied, as described below. 
Finally, we impose $H_T$ and invariant mass cuts on all events and combine the three categories in a joint likelihood.

\paragraph{Fully separated category:}
For $m_S\lesssim\SI{1.5}{\TeV}$, for most signal events the four $b$-jets are fully separated at truth level, so that they can be treated as four individually reconstructed jets.
As discussed above the background processes relevant for this case are the \ac{VBF} $hh$, $Zh$ and $ZZ$ backgrounds, as well as the continuum backgrounds $t\bar{t}h$ and $\gamma\gamma\to hh$.
For signal and background events where the four $b$-jets are fully separated we require that both jets associated with a Higgs boson are tagged as $b$-jets.
We further require both jet pairs to satisfy
\begin{equation}\label{eq:m2bcut}
    |m_{bb}^{(i)} - m_h| < 15\,\text{GeV} \quad \text{for}\; i=1,2.
\end{equation}

\paragraph{Fully merged category:}
If both $b$-jet pairs have a separation of $\Delta R_{bb} < 0.4$, as is the case for most events for $m_S\ge \SI{2.5}{\TeV}$, we assume that each pair is reconstructed as a single jet with two displaced $B$-meson vertices.
The fake double-$b$ jet backgrounds $c\bar{c}$, $b\bar{b}$ and $t\bar{t}$ must be included, in addition to the $hh$, $Zh$, $ZZ$, $t\bar{t}h$ and $\gamma\gamma\to hh$ backgrounds. 
The latter all lead to merged $b$-jets if the $h$ and/or $Z$ are highly boosted.
We assume a dedicated tagger, as described in \cref{app:tagger}, that includes a cut on the invariant mass of the jet and requires two reconstructed displaced $B$-meson vertices within the jet.
The signal efficiency for this tagger is assumed to be $\epsilon_h = 80\%$, with $b\to h$ mistag rate of $\epsilon_b=2\%$.
This is sufficient to suppress the $b\bar{b}$ background to negligible levels.
For boosted $Z\to b\bar b$ decays, we conservatively assume that their tagging rate is also $80\%$, the same as for $h\to b\bar b$. 

\begin{figure}[p]
    \centering
    \includegraphics[width=0.9\linewidth]{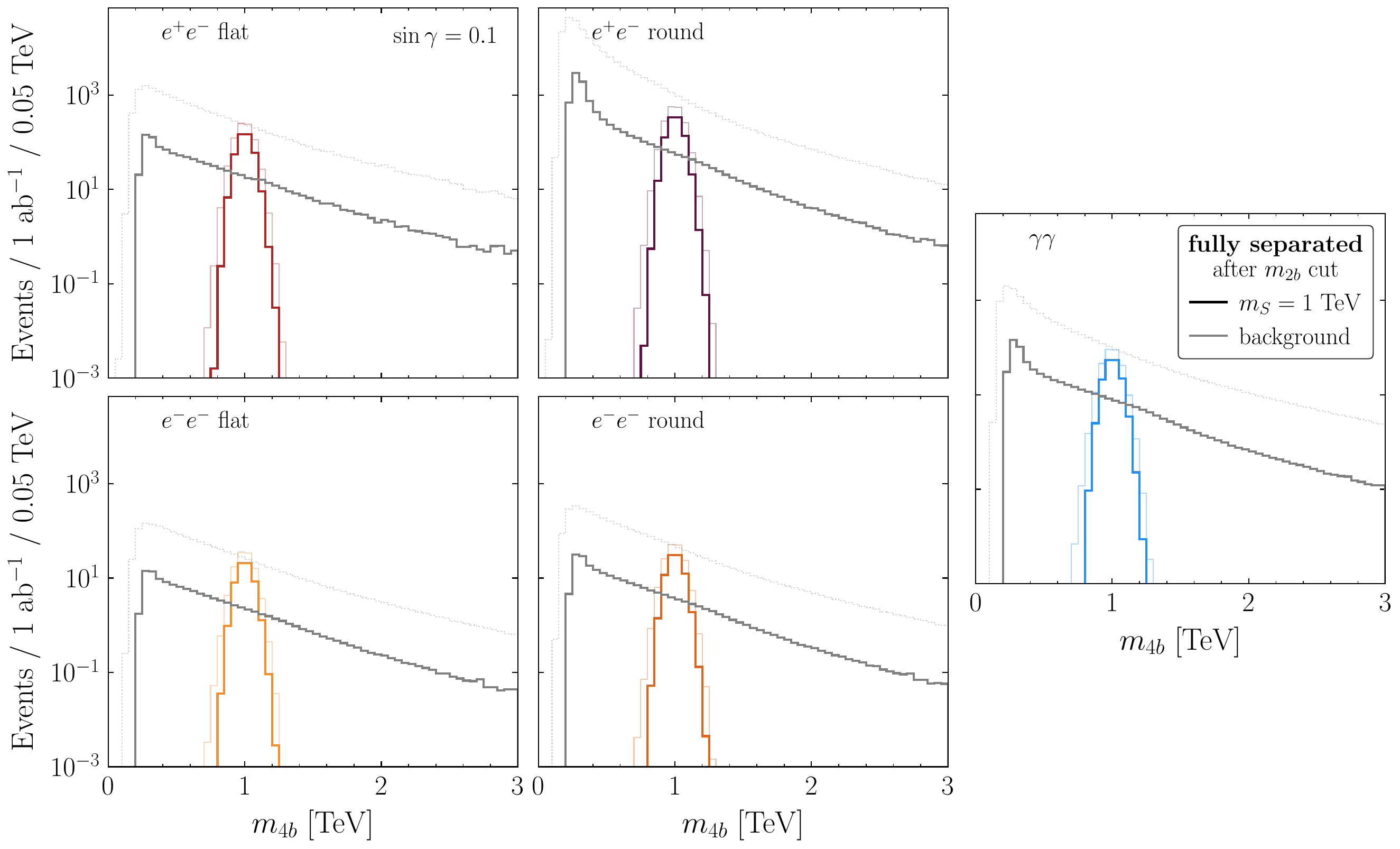}
    \caption{Invariant mass $m_{4b}$ distribution for events with fully separated jets for signal with $m_S =\SI{1}{\TeV}$ and $\sin\gamma=0.1$, and backgrounds.
Faint lines show the total number of events in this category; bold lines show the number of events after the Higgs mass window cut $|m_{bb}^{(i)} - m_h| < 15\,\text{GeV}$ and the b-tag efficiency are applied.
    }
    \label{fig:inclusive_m4b}
\end{figure}
\begin{figure}[p]
    \centering
    \includegraphics[width=0.9\linewidth]{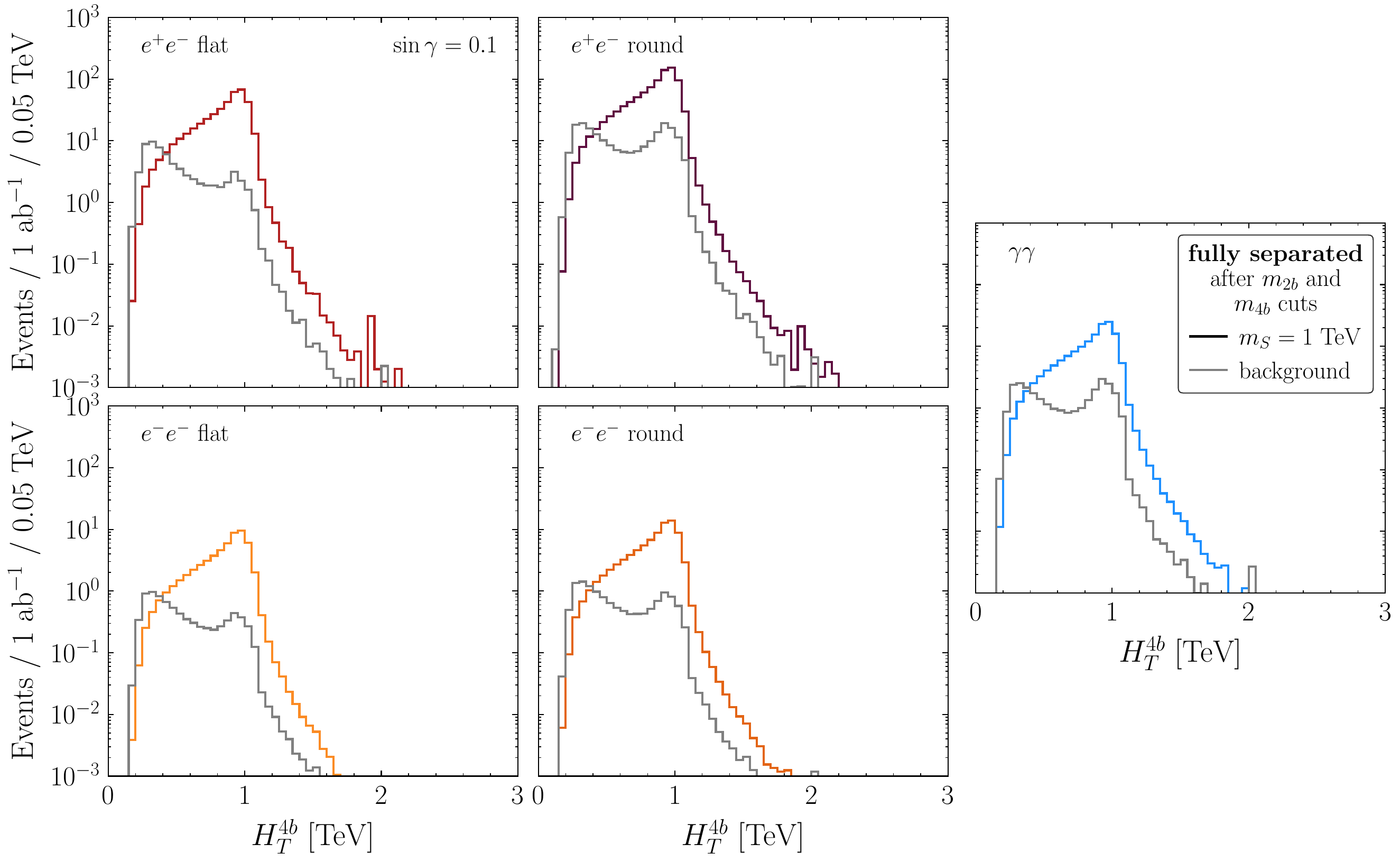}
    \caption{Scalar sum of the transverse momenta of the $b$-jets $H_T^{4b}$ of signal ($m_S =\SI{1}{\TeV}$ and $\sin\gamma=0.1$) and background events in the fully separated category after cuts on $m_{2b}$ and $m_{4b}$ have been applied.
    }
    \label{fig:inclusive_HT4b}
\end{figure}

\paragraph{Partially merged category:}
Finally, around $m_S\sim\SI{2}{\TeV}$ we find a significant fraction of events where one of the $b$-jet pairs is merged and the other is resolved.
The usual $hh$, $Zh$, $ZZ$, $t\bar{t}h$ and $\gamma\gamma\to hh$ backgrounds enter this category when one $b$-jet pair is merged and the other is resolved, so that the four $b$ quarks are reconstructed as three jets.
We explicitly checked that $4b$ backgrounds, where one $b$-jet is outside the detector coverage, are negligible.
For signal and background events in this category we require one merged tagged jet, assuming a tagger with the same signal efficiency and background suppression as in the fully merged category, while the resolved jet pair is required to consist of two b-tagged jets that fulfill $|m_{bb}^{(i)} - m_h| < 15\,\text{GeV}$.

\paragraph{All categories:}
For all categories we subsequently define the invariant mass ($m_{4b}$, \cref{fig:inclusive_m4b}) and the scalar $p_T$ sum ($H_T^{4b}$, \cref{fig:inclusive_HT4b}) of the four-$b$ system.
The invariant mass distributions of the fully separated category for signal with $m_S=\SI{1}{\TeV}$ and background are shown before and after the $m_{bb}$ cut.
They clearly indicate the expected resonance structure for the signal as compared to the continuum background.
The corresponding distributions for the fully and partially merged categories with the $m_{bb}$ cut and tagger efficiencies applied as appropriate are shown in \cref{fig:inclusive_m4b_merged} ($m_S=\SI{4}{\TeV}$) and \cref{fig:inclusive_m4b_semi} ($m_S=\SI{2}{\TeV}$) in \cref{app:figures}, showing the same resonance structure as the fully resolved scenario.

For our analysis selection, we impose 
\begin{equation}
    0.92<\frac{m_{4b}}{m_S}<1.1
\end{equation}
on the invariant mass to further reduce the background. The $H_T^{4b}$ distributions after this cut are shown in \cref{fig:inclusive_HT4b} for the fully separated category, with the remaining categories shown in \cref{app:figures} (\cref{fig:inclusive_HT4b_merged} and \cref{fig:inclusive_HT4b_semi}). 
We find that the additional selection 
\begin{equation}
0.7 < \frac{H_T^{4b}}{m_S} < 1.09 
\end{equation}
further improves signal sensitivity.

\subsection{Results}

\begin{figure}[t]
    \centering
    \includegraphics[width=0.95\linewidth]{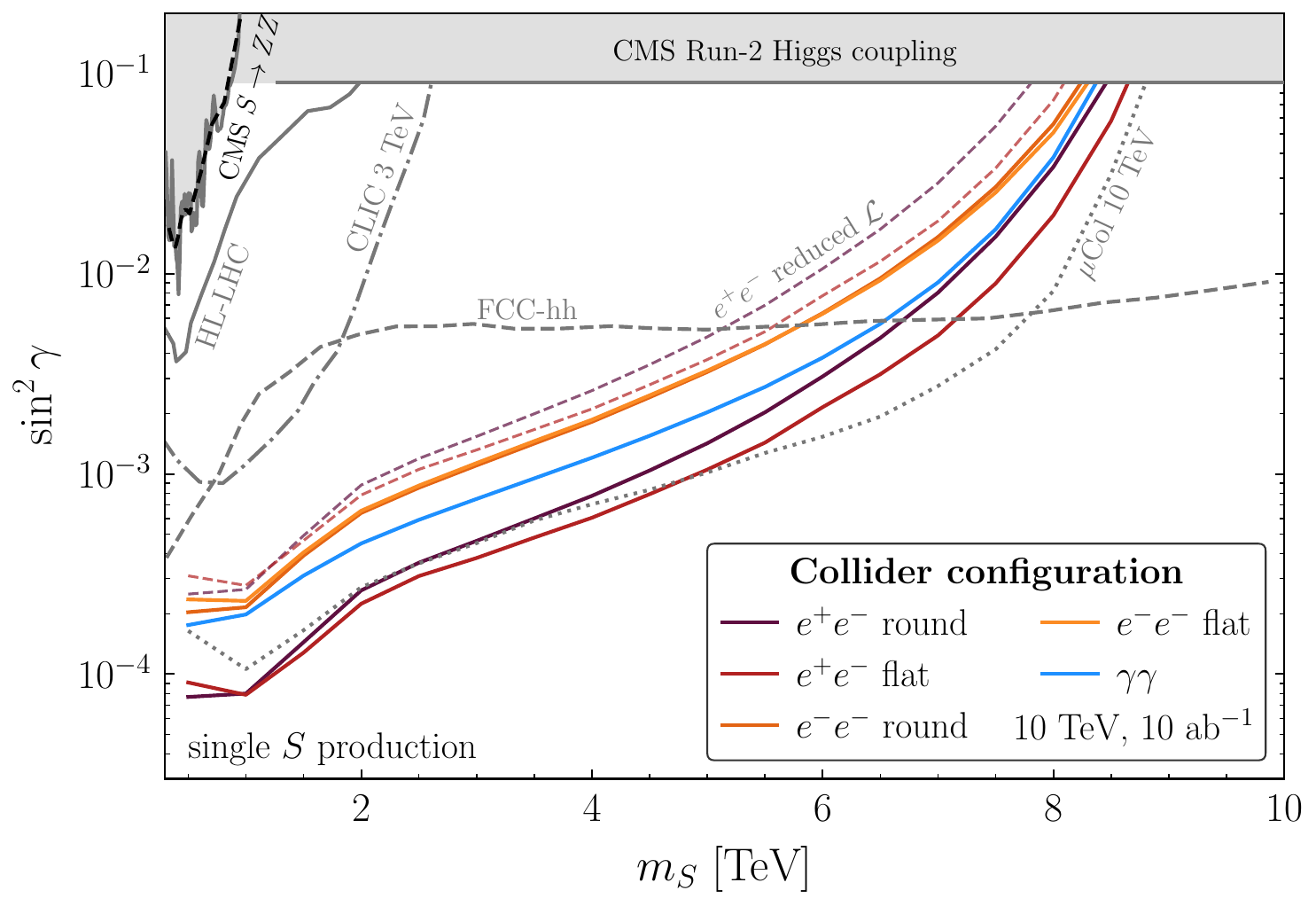}
    \caption{
        Expected $95\%$ CL upper limits on $\sin^2\gamma$ for $\lambda_{HS}=0$ and an integrated geometric luminosity of \SI{10}{\per\atto\barn} for all five wakefield collider configurations (solid) and \SI{1}{\per\atto\barn} for the $e^+e^-$ configurations (dashed).
        In addition to the projected limits for the five plasma wakefield colliders, limits from a combined measurement of the Higgs couplings at CMS \cite{CMS:2026zxu} and a CMS search for a heavy scalar resonance \cite{CMS:2026xbb} with a width of $\Gamma_S=100$~GeV (black, dashed) and in the narrow width approximation (grey solid) are shown, both using the full Run~2 data.
Also shown are projections for the HL-LHC ($\SI{3}{\rm{ab}^{-1}}$) \cite{Buttazzo:2018qqp}, CLIC ($\SI{3}{\TeV}$, $\SI{3}{\rm{ab}^{-1}}$) \cite{Buttazzo:2018qqp}, a muon collider (\SI{10}{\TeV}, $\SI{10}{\rm{ab}^{-1}}$) and FCC-hh ($\SI{100}{\TeV}$, $\SI{3}{\rm{ab}^{-1}}$) \cite{Buttazzo:2018qqp}.
   }
    \label{fig:sin2gamma_inclusive}
\end{figure}

Based on these selection cuts, for each signal mass $m_S$ we find the expected $95\%$ CL upper limit on $\sin^2\gamma$.
The statistical test is a binned log-likelihood ratio with three bins corresponding to the three jet-isolation categories (fully separated, partially merged and fully merged), assuming Poisson statistics.
The upper limits are drawn from Asimov datasets~\cite{Cowan:2010js} for each $m_S$ hypothesis, setting the observed number of events to be that predicted by the background processes alone.
We use the asymptotic approximation assuming the test statistic is $\chi^2$-distributed with a single degree of freedom.
No systematic uncertainties were included.

We show the corresponding results for all five collider configurations assuming a geometric luminosity of $\SI{10}{\per\atto\barn}$ in~\cref{fig:sin2gamma_inclusive}, though it is worth keeping in mind that these choices face very different technological challenges, as reviewed in the introduction.
In particular, acceleration of high luminosity $e^+$ beams is qualitatively more difficult than that of $e^-$ beams, which may reduce the achievable luminosity of $e^+e^-$ machines by an order of magnitude or more compared to $e^- e^-$ or $\gamma\gamma$ counterparts.
To estimate this effect, the expected reach of an $e^+e^-$ machine with a reduced geometric luminosity of $\SI{1}{\rm{ab}^{-1}}$ is shown in dashed lines; if this is deemed to be the more fair comparative benchmark, then the performance is led by the $\gamma\gamma$ collider. 
 
To quantify the impact of the beam-beam interactions, we performed our analysis also without including beam-beam interactions, with identical analysis cuts and a luminosity of $\SI{10}{\per\atto\barn}$. 
This setup is also a good proxy for a muon collider, and is labeled as such in \cref{fig:sin2gamma_inclusive}. 
We see that it yields slightly different sensitivity from the $e^+e^-$ wakefield colliders; this difference in sensitivity is entirely due to the luminosity smearing through the beam-beam interactions.
 Owing to the concentrated luminosity at peak energies, the muon collider performs better for higher mass singlets and slightly worse for lower mass ones. 
The reason is that the dependence of the cross section is logarithmic in the center-of-mass energy $M$, which favors a collider with as much luminosity as possible at high energies.
This also partially explains the slight advantage of flat $e^+e^-$ beams over round $e^+e^-$ beams: the energy scaling of the production cross section compensates for the overall higher collisional luminosity in round beams, resulting in approximately similar signal and \ac{VBF}-induced background yields between round and flat beam colliders over much of the mass range. On the other hand, the round beams produce significantly more background events from non-\ac{VBF} processes and end up slightly less sensitive than flat beams after all cuts are applied.
This is in contrast with pair production of charged particles, for which the cross section scales as $\sim 1/M^2$ above the production threshold, thus generally favoring a broader luminosity spectrum if the mass of the new particles is not a priori known \cite{Chigusa:2025azs}.
The reach for FCC-hh is taken from~\cite{Buttazzo:2018qqp}.
This was not intended as a first principles study of the FCC-hh capabilities, however, as it rescales an existing $\SI{13}{\rm{TeV}}$,  $\SI{36}{\rm{fb}^{-1}}$  CMS search~\cite{CMS:2018amk} with the ratios of the parton luminosity functions to its expected performance at $\SI{100}{\rm{TeV}}$,  $\SI{3}{\rm{ab}^{-1}}$.
FCC-hh sensitivity to heavy singlet scalars has also been studied recently in the real singlet model in~\cite{Bosse:2026bdk} and in the complex singlet model in~\cite{Lane:2024vur}.

\section{Pair production\label{sec:singletpairprod}}
\begin{figure}
    \centering
    \includegraphics[width=1.0\linewidth]{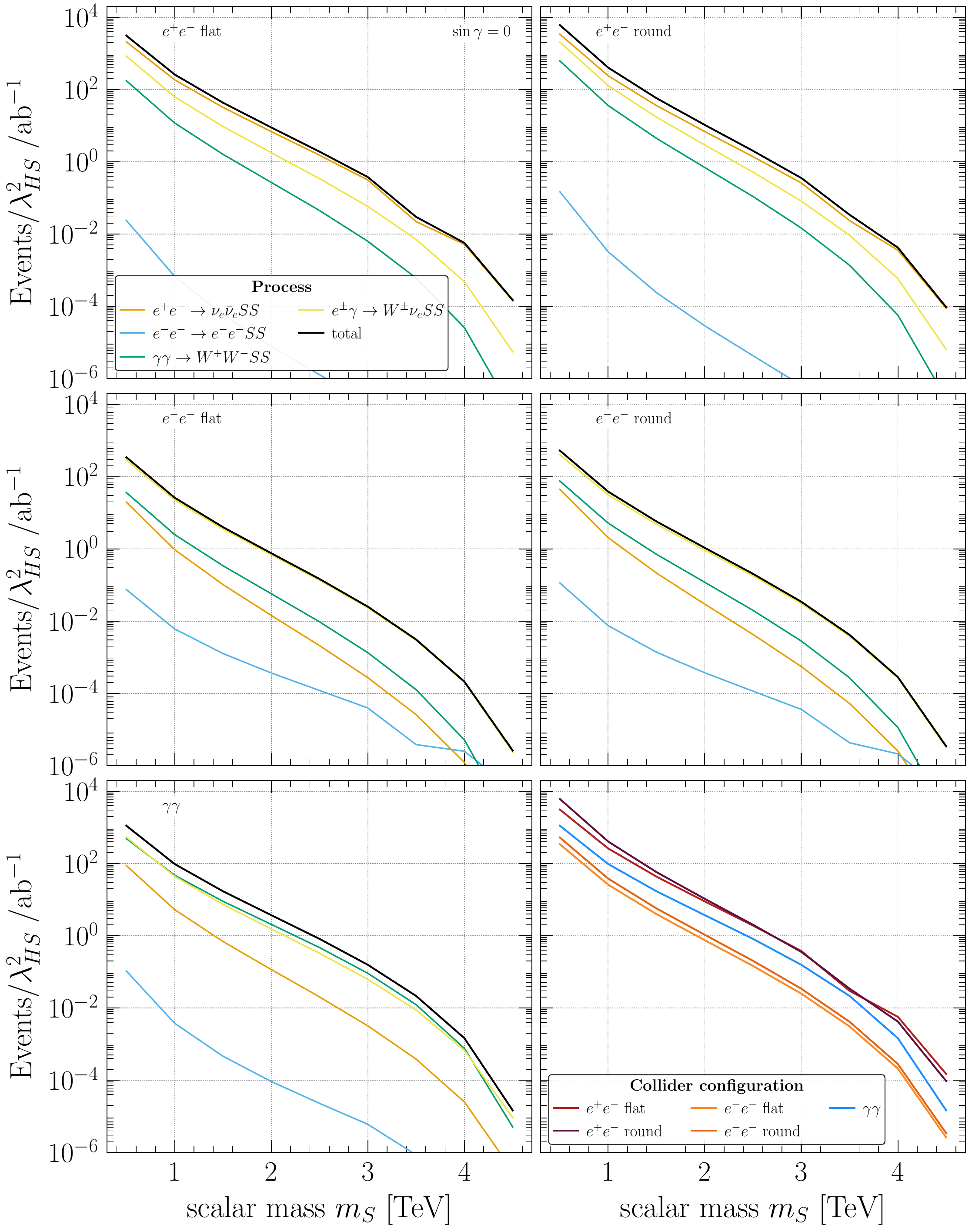}
    \caption{Number of inclusive pair-production events as a function of the scalar mass,  assuming $\sin\gamma=0$ and production only via $\lambda_{HS}$. }
    \label{fig:double_events}
\end{figure}

For $\lambda_{HS}\sim \mathcal{O}(1)$ and $\sin\gamma\ll1$, the rate of pair production dominates over that of single $S$ production.
We study pair production via four channels: $e^+e^-\rightarrow SS \nu \bar{\nu}$, $e^- e^- \rightarrow SS e^- e^-$, $e^{\pm} \gamma \rightarrow SS W^\pm \nu $ and $\gamma\gamma \rightarrow SS W^+ W^-$. 
The overall number of scalar pairs produced is obtained by folding the cross sections with the collider's luminosity functions, and is shown in \cref{fig:double_events}. 
As for the single production case, processes relying on the $W$ coupling to the electrons are the dominant production channels in all collider configurations: The $e^+e^-$ initial state dominates at $e^+e^-$ colliders as opposed to the $e^-\gamma$ initial state at $e^-e^-$ colliders.
For the $\gamma\gamma$ collider both the $\gamma\gamma$ and $e^-\gamma$ initial states contribute comparably.

\subsection{Prompt decays\label{sec:pairprompt}}
 First, we will assume that $\sin\gamma$ is too small to facilitate meaningful rates in single production, but 
 large enough for the $S$ to decay promptly compared to the detector's vertexing resolution.  
The signature of a pair of heavy $S$ particles decaying to $W$, $Z$ and $h$ bosons is expected to be spectacular in this case, with a relatively low \ac{SM} background.
For simplicity, here we only consider the fully hadronic final states, where all bosons decay to a pair of jets. Specifically, we studied:
\begin{itemize}
    \item The 8 jet final state without $b$-tags, where the $SS$ pair is allowed to decay to any combination of $WW$, $ZZ$ and $hh$ pairs. This channel has the highest overall signal yields of all channels, but we find a non-negligible background from the $\gamma\gamma \to 4W+Z$ process in the \ac{SM}. Accounting for the hadronic branching ratios of the $W$, $Z$ and $h$, the branching ratio of $SS\to 8$ jets is between $16\%$ and $22\%$, depending on $m_S$. 
    \item The 4 jets + 4 $b$-jets final state, where at least one $S$ must decay to an $hh$ pair, for which we consider the dominant 4 $b$-jets branching ratio. 
    The branching ratio for the 4 jets + 4 $b$-jets final state is about $7\%$.
    The signal yield is therefore smaller than for the 8 jet final state, but we predict negligible backgrounds for the luminosities we consider.
\end{itemize}
The channel with 4 jets and 4 $b$-jets gives slightly better sensitivity than the 8-jet channel, so we focus on it in this section and discuss the 8-jet channel in \cref{app:8jetsbackground}.
There is however a caveat worth mentioning:
For $\sin\gamma\ll1$, the destructive interference in \cref{eq:kappa-h-general} causes the tree-level $S\to hh$ branching ratio to vanish around \begin{equation}
  \lambda_{HS}\approx\frac{m_S^2+2m_h^2}{2v^2}.  
\end{equation} This suppresses the signal rate in the 4 jet $+$ 4 $b$-jet channel in this isolated region of the parameter space.
However, we expect that this loss of sensitivity can largely be recovered by considering other channels, such as the 8 jet channel. We did not study this in detail in this work.

Because the decay chains contain no hard neutrinos, we can in principle reconstruct them fully, up to the jet energy resolution of the detector.
We require the four jets from the $hh$ decays to be $b$-tagged, assuming a tagging efficiency of $90\%$, and impose no tagging requirement on the remaining four jets.
The energies of the jets are smeared as in \cref{sec:detector}.

 First, we require all eight jets to lie within the detector acceptance, $|\eta|\leq 3$.
This requirement is highly efficient, with only a mild loss for scalars produced with low transverse momentum in collisions at low \ac{COM} energies, below roughly \SI{2}{\TeV}.
These collisions are often asymmetric in the energies of the incoming particles, and the resulting longitudinal boost can push the jets from one or more of the bosons outside the acceptance.
We then require each accepted jet to have $p^j_T\geq\SI{30}{\GeV}$. This cut is more restrictive, reducing the signal cumulative efficiency to about $36\%$ for $m_S=\SI{0.5}{\TeV}$ and $84\%$ for $m_S=\SI{4.5}{\TeV}$.
The next step is to reconstruct the bosons that produced the jets, demanding two Higgs bosons from two pairs of $b$-jets and two further $h$, $W$, or $Z$ bosons from the remaining jets. We define a successfully reconstructed \ac{EW} boson by demanding that the dijet invariant mass satisfies $|1-m_{jj}/m_{\rm EW}|<0.12$, where $m_{\rm EW}$ is the mass of the corresponding boson ($m_W$, $m_Z$, or $m_h$). 
Because the decay products of $S$ are typically boosted, this pairing mostly yields the correct jet assignment.
Separating $W$ from $Z$ bosons reliably requires very good jet energy resolution and may be feasible at a lepton collider (see e.g.~\cite{CLICdp:2018vnx}).
However, to be conservative, we do not assume that the detector can distinguish dijet resonances from $W$ and $Z$ decays.

Next, for a given $m_S$ hypothesis, we require that the four reconstructed bosons can be grouped into two pairs whose invariant masses both lie within $10\%$ of $m_S$.
Again, this selection is highly efficient, retaining about $85\%$ of the signal.
Because every signal topology includes some combination of forward $W$ bosons, neutrinos, and electrons, the $SS$ system recoils against these particles and acquires a small but nonzero transverse momentum.
We therefore require the vector sum of the transverse momenta of the jets, $\big|\sum_j \vec{p}_{T,j}\big|$, to exceed \SI{50}{\GeV}, which eliminates backgrounds without such recoil, such as $\gamma\gamma\to W^+W^-hh$.

For heavier scalars, the $W$, $Z$, and $h$ bosons can be boosted enough that their two daughter jets merge into a single jet. As in the single-production analysis, we consider the daughter jets of a boson merged when $\Delta R_{jj}<0.4$.
We find that scalars with $m_S\lesssim\SI{0.5}{\TeV}$ mostly produce fully resolved jets. Typically, at most one jet pair is merged for $\SI{0.5}{\TeV}\lesssim m_S\lesssim\SI{1}{\TeV}$, at most two for $\SI{1}{\TeV}\lesssim m_S\lesssim\SI{2}{\TeV}$, and three or more for $m_S\gtrsim\SI{2.5}{\TeV}$.
To identify bosons whose daughter jets have merged, we assume an ``\ac{EW} boson tagger'' based on jet mass and substructure variables, along the lines of~\cite{Bhattacharyya:2025kvl}. We take this tagger to have a signal efficiency of $90\%$ and to mistag $10\%$ of jets that do not originate from an \ac{EW} boson.

\begin{figure}
    \centering
    \includegraphics[width=0.8\linewidth]{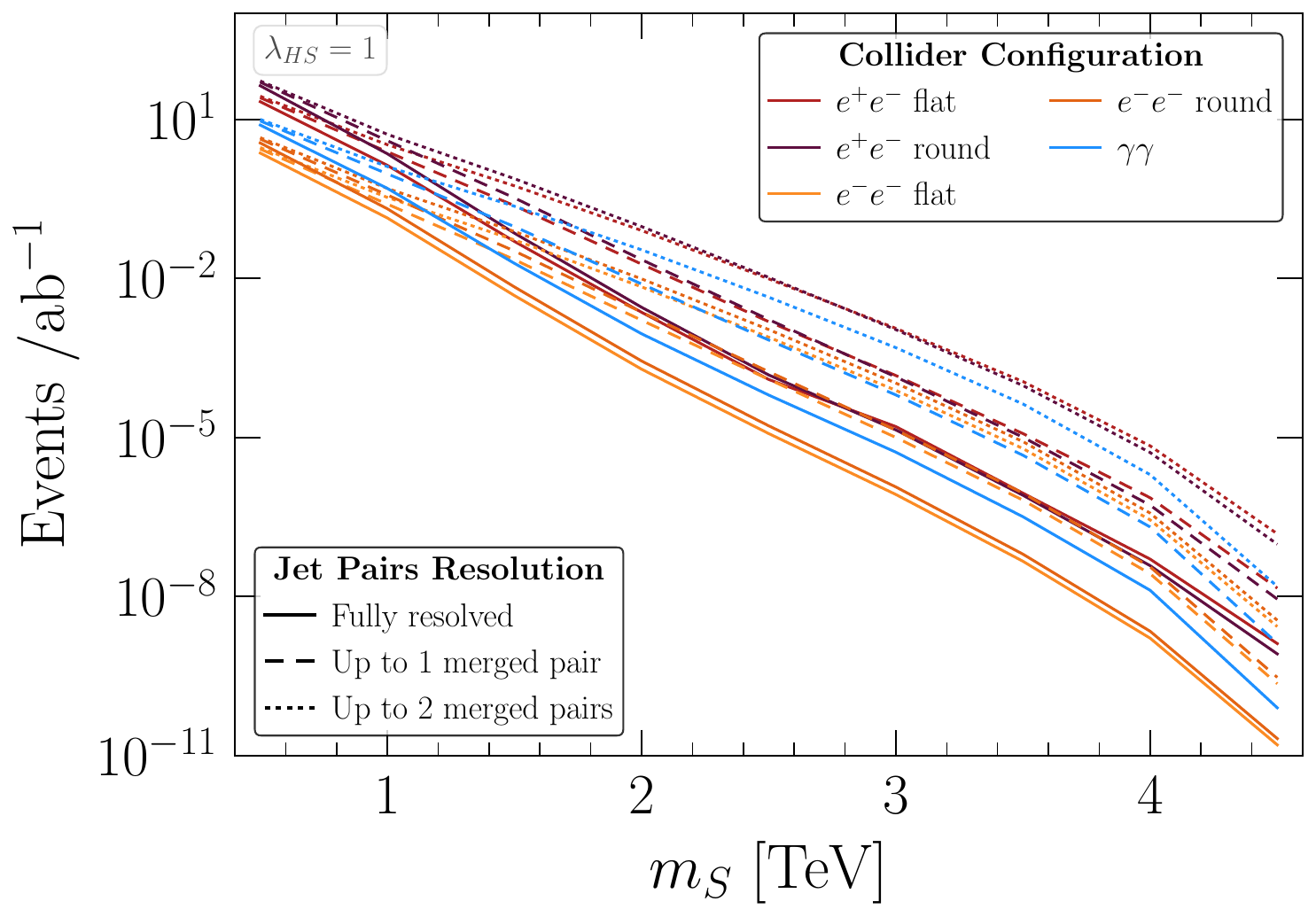}
    \caption{Number of signal events passing the kinematical cuts at an integrated luminosity of 1 $\text{ab}^{-1}$ at $\lambda_{HS}=1$.}
    \label{fig:double_events_after_cuts}
\end{figure}

A summary of the cuts applied is shown in Table~\ref{tab:cuts_double} of \cref{app:pairprod}, and the cumulative signal efficiencies for these selections are summarized in Table~\ref{tab:cutflow_eff} of \cref{app:pairprod}.
\Cref{fig:double_events_after_cuts} shows the number of signal events passing all cuts for $1\,\text{ab}^{-1}$ of integrated luminosity. 
After these selection cuts, we find an event yield for the main background processes $e^+e^- \to W^+W^-hh\nu\bar\nu$ and $\gamma\gamma \to W^+W^-Zhh$ 
that is less than $1$ in all colliders, assuming a geometric luminosity of $10  \,\text{ab}^{-1}$. We will therefore treat this channel as background free.

\begin{figure}[t]
    \centering
    \includegraphics[width=1\linewidth]{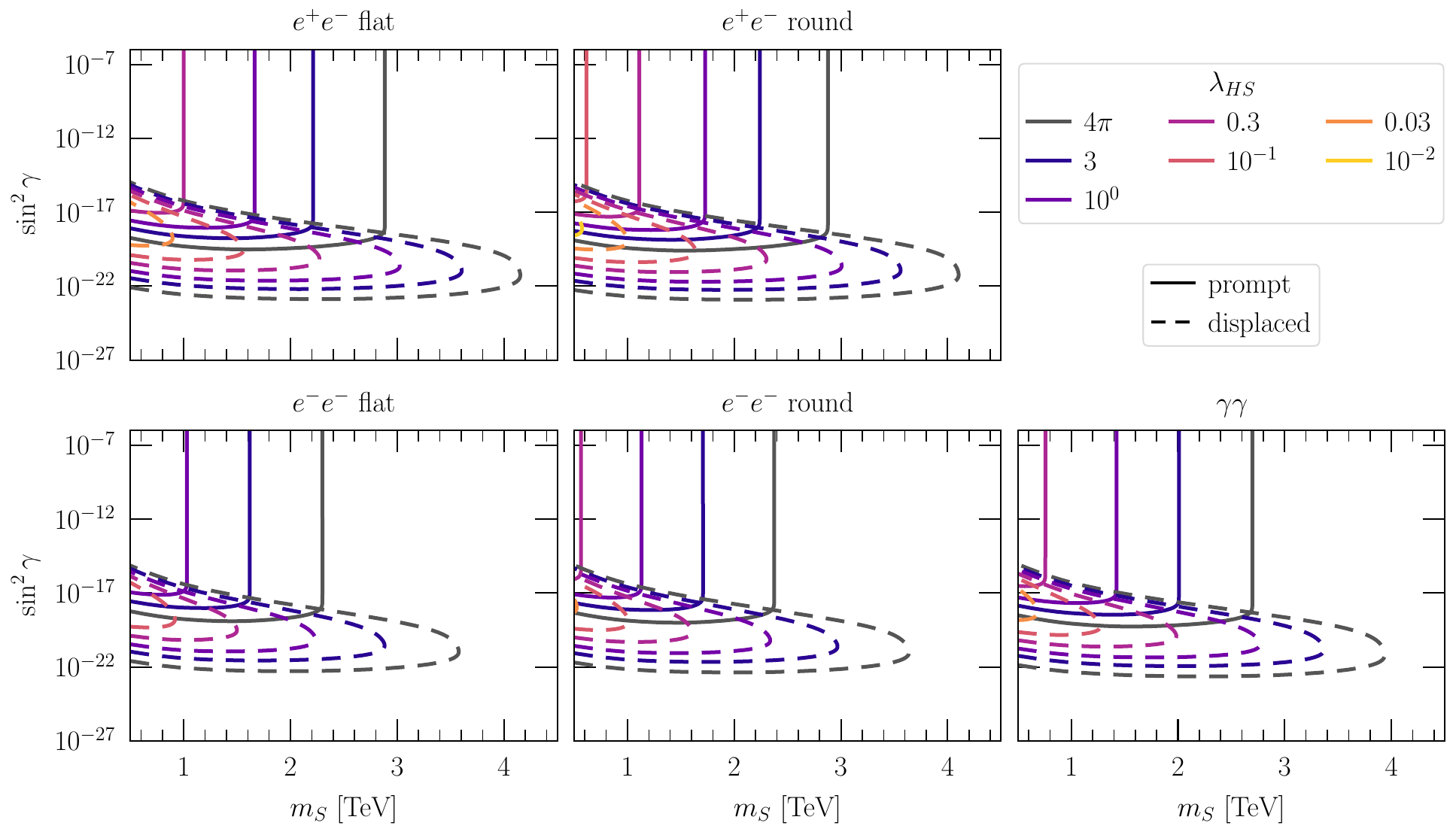}
    \caption{Expected bounds from pair-produced scalars with prompt (solid) and displaced (dashed) decays for $10~{\rm{ab}}^{-1}$ integrated luminosity. For $\lambda_{HS}\approx 4\pi$, there is a region around $m_S\approx 1$ TeV for which the $S\to hh$ branching ratio drops to zero, weakening the sensitivity of the prompt 4 $b$ + 4 jets analysis. This is not shown on the figure; we expect the 8 jets channel to dominate the sensitivity in this region. See \cref{sec:pairprompt} for details.  }
    \label{fig:bounds_double}

\end{figure}

\subsection{Displaced decays}
The signature becomes even more striking if $S$ is long-lived enough for the decay to be reconstructed as displaced. 
If hard prompt tracks can be vetoed effectively by the detector, there is no \ac{SM} background for a pair of very energetic, displaced jets: The decay products of long-lived \ac{SM} mesons are expected to form a jet together with prompt activity from the parton shower. 
The same is true for secondary vertices induced by scatterings of long-lived, neutral hadrons in the detector.
Finally, neutrinos would evade such a veto, since they leave no prompt activity, but too few energetic ones traverse the central detector to scatter in the calorimeter at an observable rate.
For the purpose of our estimates, we will therefore assume that the displaced channel is also background-free for all colliders.

For an event to be accepted, we require that both scalars decay at a transverse distance $R_T$ of \SI{5}{\cm}--\SI{10}{\m} away from the interaction point, and that at least one of the jets produced has $p_T\geq \SI{30}{\GeV}$. The branching ratio of the $SS$ system to at least one jet is very close to 100\%.

\subsection{Results}

\Cref{fig:bounds_double} shows the expected 95\% CL upper limit contours in the $\sin^2\gamma$-$m_S$ plane for various values of $\lambda_{HS}$ for an integrated luminosity of $\SI{10}{\per\atto\barn}$.
Since both the prompt and the displaced categories are expected to be background-free after appropriate cuts are applied, these bounds correspond to 3 signal events, the 95\% CL Poisson upper limit for zero observed events.
In contrast to the single production case, round beam colliders perform noticeably better than their flat beam counterparts at low scalar masses, while performing approximately similarly at larger ones.
While the trend is the same, the overall performance slightly favors round beams over the range of accessible masses.
This is however a very mild effect overall.
Similar to the single production case, the physics reach is led by $e^+e^-$ machines only if equal luminosity is achievable with all collider options.

\section{Discussion\label{sec:conclusions}}

\begin{figure}
    \centering
    \includegraphics[width=0.9\linewidth]{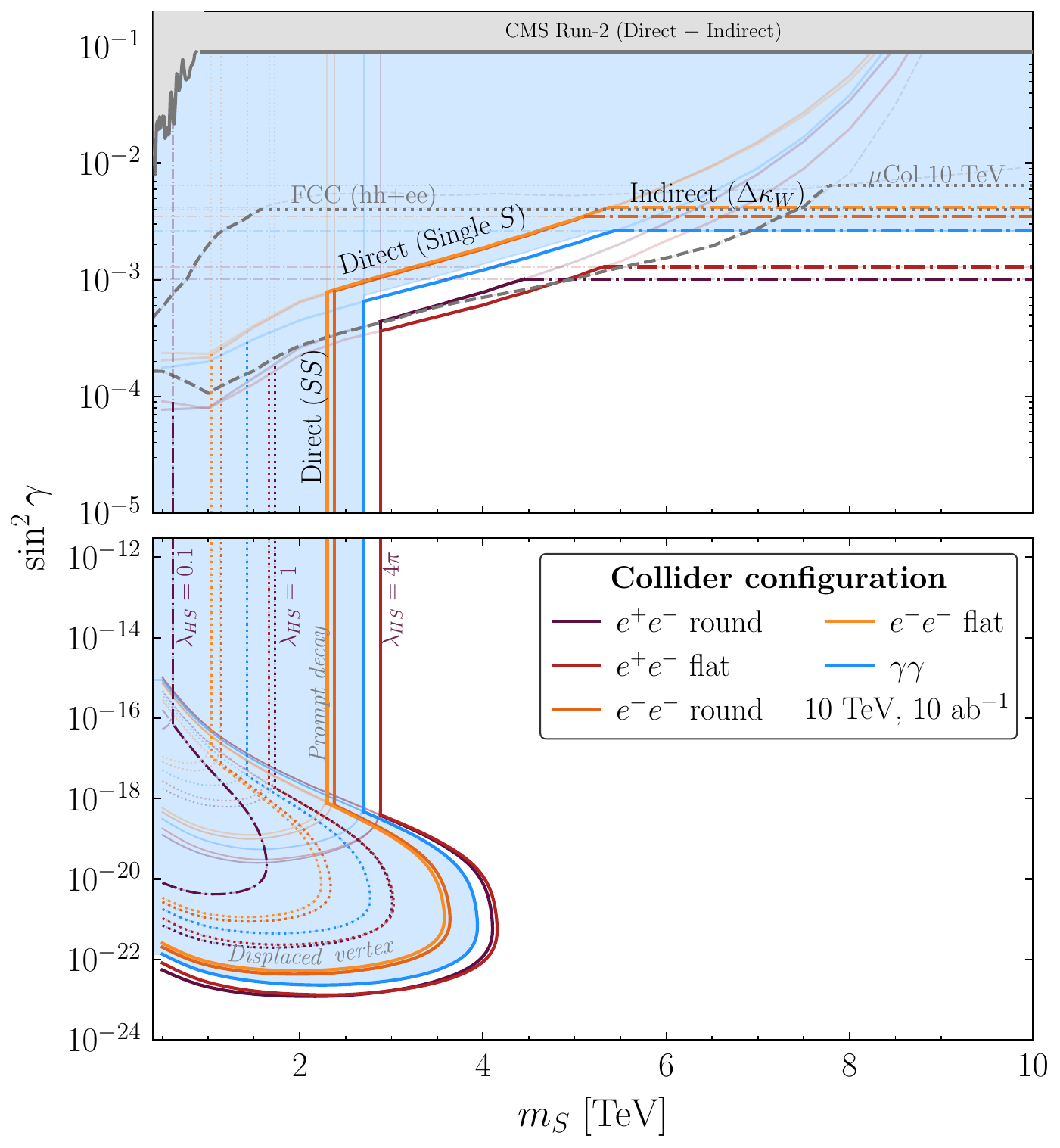}
    \caption{Regions of the scalar singlet parameter space where the sensitivity of a wakefield collider experiment is led by single production searches, pair production searches, and precision Higgs measurements respectively.
All lines represent constraining power at 95\% CL.
For illustration, the parameter region that is expected to be probed with a 10 TeV $\gamma\gamma$ collider is shaded in blue.
Note that the parameter region covered by the pair production search is set by the quartic $\lambda_{HS}$, which needs to be quite large for appreciable sensitivity.
The projected indirect sensitivities were taken from~\cite{Fraser:2026ngs}, which studied Higgs physics at a future wakefield facility.
Also shown for comparison is the expected reach (single production and Higgs precision only) of comparable future facilities.
For FCC the direct reach (dashed) assumes the hh configuration ($\SI{100}{\TeV}$, $\SI{3}{\per\atto\barn}$)~\cite{Buttazzo:2018qqp} and the indirect reach (dotted) assumes ee ($\SI{365}{\GeV}$, $\SI{3.1}{\per\atto\barn}$)~\cite{FCC:2025lpp,deBlas:2026vrz} and is driven by constraints on $\kappa_Z$.
For the muon collider ($\SI{10}{\TeV}$, $\SI{10}{\per\atto\barn}$) the direct reach is from this work, and the indirect reach is taken from~\cite{Forslund:2022xjq}.
All indirect constraints shown here assume negligible contributions from $\lambda_{HS}$.
    }
    \label{fig:singlet_space}
\end{figure}
We combine the single production, pair production and indirect probes altogether in \cref{fig:singlet_space}.
For the values of $\sin^2\gamma$ displayed in the top panel, the scalar decays promptly, and the pair-production reach is approximately independent of the mixing angle for fixed branching fractions.\footnote{The displayed reach assumes $\mathrm{Br}(S\to hh)=1/4$ for simplicity.
There are isolated regions in parameter space where this assumption is not valid; see the end of \cref{sec:pairprompt}.
In those cases the 8-jet channel is expected to perform better.}
The $\lambda_{HS}$ that is needed for pair production to be relevant is generally quite large for $m_S\gtrsim 1$ TeV, approaching unitarity saturation. 
For heavier singlet masses, direct production is suppressed, and the parameter space is more effectively probed via precision measurements of the $hWW$ coupling, as estimated in \cite{Fraser:2026ngs}. (These estimates did not include systematic uncertainties yet, and may therefore be somewhat optimistic.)
At tree level, the single-Higgs gauge-coupling shift $\Delta\kappa_V=\cos\gamma-1$ is independent of $m_S$ at fixed mixing.\footnote{Holding $\gamma$ fixed as $m_S$ increases requires $a_{HS}\propto(m_S^2-m_h^2)\sin\gamma\cos\gamma/v$ to grow. At fixed $a_{HS}$, the mixing contribution decouples.} 
Loop corrections involving $\lambda_{HS}$ introduce an additional dependence on $m_S$ (see e.g.~\cite{Craig:2013xia,Haisch:2020ahr});  in \cref{fig:singlet_space} we however fix $\lambda_{HS}=0$ when considering the indirect constraints. 

At smaller values of $\sin^2\gamma$ single production ceases to be efficient. Here the pair production governed by $\lambda_{HS}$ dominates the event yield, if $\lambda_{HS}$ is large enough. This is shown by the vertical lines in the upper panel of \cref{fig:singlet_space} for several values of $\lambda_{HS}$.
For the small mixing angles shown in the bottom panel of \cref{fig:singlet_space}, the singlet becomes long-lived enough to produce displaced vertices, providing a distinctive experimental signature.  
This means that all decay modes are essentially background free, and unlike for the prompt case, where we focused on the four jets + four $b$-jets final state (see \cref{sec:singletpairprod}) all hadronic final states are included.
This explains the improved sensitivity in this region.
At still smaller mixing angles, the singlets escape the detector before decaying. Direct searches would then rely on missing-momentum signatures with a visible recoil, which we do not study here.

Across the search modes studied here, the $e^+e^-$ configurations provide the largest signal yields and the strongest sensitivity at equal geometric luminosity. 
A $\gamma\gamma$ collider is however an attractive alternative if delivering high-energy positron beams reduces the attainable geometric luminosity by roughly a factor of three or more relative to the $\gamma\gamma$ option. 
The differences between flat and round beams are more modest: Flat beams are slightly favored, owing partially to the logarithmic energy-dependence of the production cross sections.

This is significant to the broader comparison between these collider options, because our benchmark model was specifically chosen to extend and complement prior work on direct searches for charged particles \cite{Chigusa:2025azs} and heavy resonances \cite{Cipressi:2026wtb}.
This is because its dominant production modes all relate to $W/Z$ boson fusion, as opposed to $e^+e^-$ and/or $\gamma\gamma$ annihilation in the latter examples (see Fig.~1 of Ref.~\cite{Cipressi:2026wtb} for a comparison of
the energy dependence of fusion and annihilation cross sections).
In those prior benchmarks, we found that beam-beam interactions \emph{enhance} the sensitivity, thus favoring round beams. 
Here we find that this conclusion is reversed for signatures relying on vector boson fusion. This difference is the result of several confounding factors: the larger luminosity tail induced by beam-beam interactions still results in overall higher event rates for both signal and background in much of the mass range. However, because the production cross section rises logarithmically with the center-of-mass energy, the amount of signal gained is weakened relative to the background, and final analysis sensitivity somewhat favors flat $e^+e^-$ beams over round $e^+e^-$ for most of the masses we studied.
In the end, the difference is fairly mild.

We also compare our projected sensitivities with those of the FCC-hh and of a muon collider operating at the same energy and integrated luminosity.
Since the muon collider's luminosity is concentrated near the nominal collision energy, it has better reach for heavier singlets and slightly worse reach for lighter ones than the $e^+e^-$ wakefield configurations.
This is again because of the logarithmic growth of the \ac{VBF} cross section with the center-of-mass energy $M$. 
The FCC-hh can reach the highest $m_S$ through a direct search, though precision measurements of the $hWW$ or $hZZ$ coupling at the FCC-ee \cite{FCC:2025lpp,deBlas:2026vrz}, muon \cite{Han:2020pif,Forslund:2022xjq} or wakefield colliders \cite{Fraser:2026ngs} have the potential to probe similar parameter space.

Our conclusions can be summarized as follows:
\begin{itemize}
    \item For the $e^+e^-$ colliders, beam-beam interactions indeed degrade the sensitivity at high $m_S$, but slightly improve it at low $m_S$. This is in contrast with the models studied in \cite{Chigusa:2025azs,Cipressi:2026wtb}, for which beam-beam interactions ended up strictly enhancing the sensitivity.
    These are however $\mathcal{O}(1)$ effects, and may ultimately end up being less relevant than the differences between the best achievable luminosities for round and flat beams. 
    \item For the same geometric luminosity, $e^+e^-$ colliders outperform the $\gamma\gamma$ and $e^-e^-$ options. However if the maximal achievable luminosity in $e^+e^-$ ends up being a factor of three or more smaller than that of a $\gamma\gamma$ collider, the latter provides comparable or better sensitivity (see \cref{fig:sin2gamma_inclusive}). 
    \item As for a muon collider and the FCC-ee/FCC-hh combination, we find strong complementarity between direct searches for an extended Higgs sector and precision measurements of the Higgs couplings.
\end{itemize}

\acknowledgments
The authors gratefully acknowledge the essential contributions of collaborators within the LBNL Accelerator Technology \& Applied Physics Division and the LBNL ATLAS group. Special thanks are owed to Stepan Bulanov, Arianna Formenti, Remi Lehe, Jens Osterhoff, Carl Schroeder, and Jean-Luc Vay, who generously shared their preliminary findings on accelerator parameters and beam–beam simulations. The authors also wish to thank Simone Pagan Griso and Angira Rastogi for their valuable contributions regarding detector-related assumptions. The authors are likewise grateful to all of the individuals named above for their thoughtful feedback on the manuscript and for the ongoing discussions that helped sharpen key concepts and guide the direction of this work. Thanks are also extended to Daniel Downey for his assistance in advancing a publicly accessible version of the analysis code.

The authors further wish to express their appreciation to Wolfgang Altmannshofer, Tim Barklow, Innes Bigaran, Massimo Cipressi, Nathaniel Craig, Katie Fraser, Spencer Gessner, Kevin Langhoff, Zoltan Ligeti, David Marzocca, Pankaj Munbodh, Michael Peskin, Niki\v{s}a~Ple\v{s}ec, Nick Rodd, Robert Szafron, and Jure Zupan for their helpful remarks, exchanges, and/or joint work on related topics.

Financial support for SC was provided by the Simons Foundation through Investigator grant 929241, as well as by the U.S. Department of Energy, Office of Science, through the BNL C2QA program under grant Contract Number DE-SC0012704 (SUBK\#390034).
The research of SK, CS, and IS is supported by the Office of High Energy Physics of the U.S. Department of Energy under contract DE-AC02-05CH11231.
IS additionally acknowledges support from the Leinweber Foundation and from the Weizmann Institute of Science Women's Postdoctoral Career Development Award.
WLX received support from the Kavli Institute for Particle Astrophysics and Cosmology and from the U.S. Department of Energy under contract DE-AC02-76SF00515.
Parts of this work were carried out at the Aspen Center for Physics, which receives support from the National Science Foundation under grant PHY-2210452.
Computational resources were provided by the National Energy Research Scientific Computing Center (NERSC), a U.S. Department of Energy Office of Science User Facility, under NERSC award HEP-ERCAP0031191.

\appendix
\crefalias{section}{appendix}
\crefalias{subsection}{subappendix}
\crefalias{subsubsection}{subsubappendix}

\section{Comments on exclusive analysis strategies}
\label{sec:exclusive}

\subsubsection*{\boldsymbol{$W^+W^-S$}}
For a $\gamma\gamma$ collider an exclusive search could be based on $\gamma\gamma\to W^+W^- S$ where the $W$ bosons decay hadronically.
Then, the signature would be four $b$-jets with four additional jets.
For this signature fewer background processes than for an inclusive analysis contribute.
Specifically, we find that $8j$ backgrounds are small, so that only $W^+W^-hh$, $W^+W^-Zh$, $W^+W^-ZZ$ and $e^- \gamma \to \nu_e W^- Z h h$ need to be taken into account.
At the same time there will be about a factor of two fewer signal events compared to an inclusive analysis, because the contribution from $e^\pm \gamma \to W^\pm \nu_e S$ is no longer considered.
Applying similar cuts as discussed in \cref{sec:singletanalysiscuts} we find that an exclusive $W^+W^-S$ analysis always performs worse in terms of $\sin^2\gamma$ reach than the inclusive search.  

\subsubsection*{Single \boldsymbol{$\mu$}}

The single-$\mu$ channel can be viewed as an exclusive handle on the $W$-associated production modes.
Requiring an identified muon efficiently removes backgrounds initiated by an $e^+e^-$ collision, including the sizeable $e^+e^-\to \nu_e\bar{\nu}_ehh$ contribution in \cref{fig:backgroundxsec}.
However, the same requirement also removes the charged-current signal process $e^+e^-\to\nu_e\bar{\nu}_eS$, which is one of the leading production modes in \cref{fig:signal_events_inclusive_mass}.
Therefore, this channel is useful only to isolate the $e^-\gamma\to \nu_e W^- S$ and $\gamma\gamma\to W^+W^-S$ topologies and to test whether the cleaner event category selected by a single-muon tag can compensate for the loss in signal rate.

Representative signal and background rates relevant for the single-$\mu$ strategy are listed in \cref{tab:single_mu_processes}. 
Requiring $W\to\mu\nu$ multiplies the $e^-\gamma$-initiated signal and background rates by $\mathrm{Br}(W\to\mu\nu)\simeq 0.11$, while the corresponding $\gamma\gamma$-initiated rates are reduced by $2\,\mathrm{Br}(W\to\mu\nu)\left[1-\mathrm{Br}(W\to\mu\nu)\right]\simeq 0.20$.
Thus, even if one optimistically ignores the loss of signal events from the single-$\mu$ requirement, the background yield after the kinematic selection, including the Higgs-window requirement on $m_{2b}^{(i)}$, would still have to be reduced by at least a factor of about five for this strategy to be competitive with the inclusive analysis.
However, the background rates in \cref{fig:backgroundxsec,tab:single_mu_processes} differ only by factors of $\mathcal{O}(1)$ between the $e^+e^-$-initiated and non-$e^+e^-$-initiated contributions, making this difficult to achieve in practice.
Indeed, in our dedicated study of single-$\mu$ events we found a sensitivity to $\sin^2\gamma$ that is strictly weaker than the inclusive reach shown in \cref{fig:sin2gamma_inclusive}.

\begin{table}[t]
    \centering
    \small
\begin{tabular}{L{0.37\textwidth} C{0.25\textwidth} C{0.25\textwidth}}
    \toprule
      Process
      & $\sigma(\sqrt{s}=2\,\mathrm{TeV})$ [fb]
      & $\sigma(\sqrt{s}=10\,\mathrm{TeV})$ [fb] \\
    \midrule $e^-\gamma\to \nu_e W^- (S\to hh)$   & $2.5\times 10^{-3}$ & $2.8\times 10^{-2}$ \\
    $\gamma\gamma\to W^+W^-(S\to hh)$, single $\mu$ & $1.0\times 10^{-2}$ & $1.0\times 10^{-1}$ \\
    \hline $e^-\gamma\to \nu_e W^- hh$        & $1.6\times 10^{-2}$         & $1.8\times 10^{-1}$ \\
    \gray{$e^-\gamma\to \nu_e W^- Zh$}        & \gray{$5.0\times 10^{-2}$}         & \gray{$4.1\times 10^{-1}$} \\
    $\gamma\gamma\to W^+W^-hh$, single $\mu$ & $2.7\times 10^{-2}$  & $4.3\times 10^{-1}$ \\
    \gray{$\gamma\gamma\to W^+W^-Zh$, single $\mu$} & \gray{$1.3\times 10^{-1}$}  & \gray{$1.1$}  \\
    \bottomrule
\end{tabular}
    \caption{
        Cross sections of representative signal and background processes relevant to the single $\mu$ search at $\sqrt{s}=2\,\mathrm{TeV}$ and $\sqrt{s}=10\,\mathrm{TeV}$.
        The singlet mass is fixed to $m_S=1\,\mathrm{TeV}$ and the mixing angle to $\sin\gamma=0.1$.
        Entries already include the branching ratios for the $h$ and $Z$ decays, with $h/Z\to b\bar{b}$ implicitly assumed, and the branching ratio for $W^-\to \mu^-\nu$ to produce a single $\mu$.
        For background processes initiated by $\gamma\gamma$, a branching factor of $2\mathrm{Br}(W\to \mu\nu)(1-\mathrm{Br}(W\to \mu\nu)) \simeq 0.20$ is included to account for the single $\mu$ requirement.
        Gray shaded entries denote $Zh$ backgrounds that are neglected in the single-$\mu$ estimate because they become negligible after the Higgs mass window cut.
    }
    \label{tab:single_mu_processes}
\end{table}

\section{Double $b$-tagging \label{app:tagger}}

In this appendix we estimate how efficiently a merged $h\to b\bar b$ jet can be distinguished
from a single-$b$ or $c$ jet with an idealized detector at a high energy lepton-collider.
This is necessary, as hard gluon radiation in a multi-TeV $b/c$-jet is rather likely and a jet mass comparable to $m_h$ is therefore fairly typical. 
To suppress this background we must therefore also classify the jets based on the number of displaced vertices that are consistent with $B$-meson decays. 
In the background sample, such additional vertices could come from misclassified $D$-meson decays or a hard gluon splitting into an additional $b\bar b$ pair during the parton shower. 
This type of tagging algorithm could ultimately be optimized with machine learning techniques. 
For the present study, however, we find that a simple, cut-based algorithm is sufficiently powerful and has the additional advantage of being more transparent. It will also allow us to make some qualitative but clean statements about the desired detector capabilities.

\subsection{Setup}

We set up a particle gun in \textsc{Pythia}~8.317 \cite{Sjostrand:2014zea} to initialize a pair of back-to-back $h$ bosons with an energy of 2 TeV each. For the background we simulate back-to-back $b\bar b$ and $c\bar c$ pairs with the same energy. 
For all samples we run the parton shower and hadronization steps with the same settings.
Jets are clustered from all visible final-state
particles with the anti-$k_T$ algorithm \cite{Cacciari:2011ma} with a jet radius of $R=0.5$. We then select jets with
$|\cos\theta_j|<0.9$, $E_j > \SI{1}{\TeV}$ , and $100<m_{\rm jet}<\SI{150}{\GeV}$.
Since the particle-gun samples are generated at a fixed energy of $\SI{2}{\TeV}$, applying the efficiencies quoted below to jets at other energies assumes that this choice of energy is representative. We verified that doubling the energy to 4 TeV per $h$ only reduces the tagging rate by $\sim 10\%$ for the benchmark detector parameters we adopt below.

The $m_{\rm jet}$ cut retains $86\%$ of the $h\to b\bar b$ signal
jets and $17\%$ of hard background jets. This factor of $\sim 6$ rejection alone is not satisfactory given the high $b\bar b$ and $c\bar c$ cross sections, hence the need to count $B$-meson vertices.
Of the $b$ background jets in the window, $88\%$ contain a single $b$ hadron,
$5\%$ contain a $b\bar b$ pair from gluon splitting, and $7\%$ are
$b$-free gluon-radiation jets.

We do not perform track fitting but subject the truth-level decay vertices to parametric cuts that model the degree to which they could be reconstructed by a state-of-the-art tracking detector.
Concretely, a vertex is considered reconstructable if its transverse radius satisfies $r_{xy} < r_{\rm vtx}$, where $r_{\rm vtx}$ is the maximum radius at which the tracker can reliably reconstruct a displaced vertex.
We will scan over this parameter.
A vertex should further have two or more charged tracks with $p_T>0.5$~GeV.
We take the spatial vertex resolutions transverse and parallel to the decaying particle's trajectory to be $\sigma_\perp = 5\,\mu$m and $\sigma_\parallel = 20\,\mu$m respectively.
All vertices with a transverse (longitudinal) distance smaller than $3\sigma_\perp$ ($3\sigma_\parallel$) are merged into a single vertex. 
Next, measured vertices are grouped into \emph{vertex systems}: a vertex is attached to an existing system if it lies within $0.5$~mm (perpendicular distance) of the flight axis from the IP through the system's innermost vertex.
This procedure attaches downstream $D$-meson decays to their mother $B$-meson, to avoid counting them as an additional $B$-vertex.
The two $b$-quarks from an $h\to b\bar b$ decay, on the other hand, still have a sufficiently large opening angle to remain resolved separately with this criterion.
We further make the following idealized assumptions: strange hadrons are kept stable, there is no pile-up and beam-induced background and material interactions with the detector are neglected as well.

A jet is considered ``double-$b$ tagged'' if it contains $\geq 2$
vertex systems, as defined above.
In addition, we employ a variable cut on the invariant mass of the tracks in the \emph{second} system, $m_{\rm vtx,2}$, which separates a genuine second $b$ chain (vertex mass up to $\sim 5$~GeV) from a charm system ($\lesssim 2$~GeV).
This is also very effective at suppressing the $c\bar c$ backgrounds.
Varying the cut on $m_{\rm vtx,2}$ allows us to trade signal efficiency for signal purity, depending on the application at hand.
For the $b\bar b$ background sample, we find that the second system in jets with only 1 $b$-quark is almost always due to a charm quark from the parton shower ($98\%$), rather than the split-off $D$ vertex of the $b$ chain itself.
In other words, the $B$/$D$ collinearity merging is nearly fully efficient.
The residual single-$b$ fake rate is therefore mainly due to additional heavy-quark pairs from gluon splitting, rather than split-off $D$ vertices from the original $b$ chain.
The $m_{\rm vtx,2}>2$~GeV tail comes mainly from collimated $c\bar c$ pairs whose two $D$ vertices merge into one vertex.

\subsection{Results}

\begin{figure}[t]
\centering
\includegraphics[width=\textwidth]{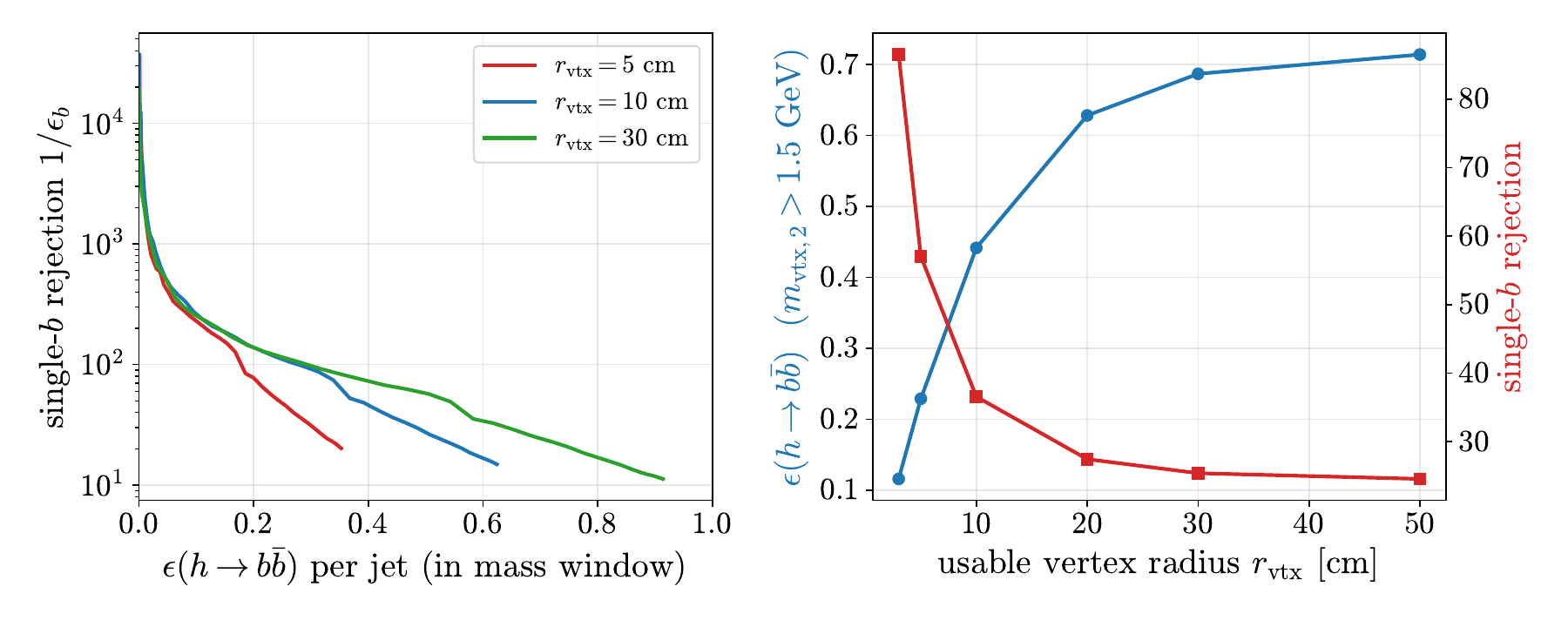}
\caption{Left: per-jet ROC curves, $h\to b\bar b$ efficiency versus single-$b$
rejection, both computed only for jets satisfying the $m_{\text{jet}}$ cut, for three values of the usable
vertex radius $r_{\rm vtx}$. Right: efficiency and rejection at the
$m_{\rm vtx,2}>1.5$~GeV working point versus $r_{\rm vtx}$.}
\label{fig:toy}
\end{figure}

With this procedure we first select the sample of signal and background events that satisfy the invariant mass cut. Within these samples we calculate the efficiencies for the $h\to b\bar b$ and single $b$ ($c$) jet to be tagged, denoted respectively by $\epsilon(h\to b\bar b)$ and $\epsilon_{1b}$ ($\epsilon_{1c}$), while varying the vertex resolutions, track $p_T$ thresholds, $r_{\rm vtx}$ and $m_{\rm vtx,2}$. 
The most important parameter is the usable vertex radius
$r_{\rm vtx}$: at $E_b\sim 1$~TeV a $B$ hadron flies
$\gamma\beta c\tau \simeq 5$--$9$~cm. One must therefore have the capability to reconstruct displaced vertices several tens of centimeters from the beam line. 
Concretely, requiring both $B$ decays
inside $r_{\rm vtx}=5$~cm caps the double-vertex efficiency at $\sim35\%$,
while $r_{\rm vtx}=30$~cm allows $91\%$ (see \cref{tab:wps} and \cref{fig:toy}).
We adopt
the aggressive choice $r_{\rm vtx}=30$~cm as the baseline and a target for future detector design.
By contrast, the result is
essentially unchanged when the vertex resolution is degraded tenfold
($5\to 50\,\mu$m), the track $p_T$ threshold is doubled, or the
system-merging distance is varied by a factor of a few.
At these energies the $B$'s are so boosted that the
vertex separations are macroscopic, and the tagger is limited primarily by the geometry of the tracking detectors rather than by spatial resolution.
Without a cut on $m_{\rm vtx,2}$ the $b$ and $c$ fake rates are comparable, with the latter slightly lower due to the requirement that the vertex has two or more charged tracks. As expected, cutting on $m_{\rm vtx,2}$ is very effective at suppressing the $c$ fake rate.

\begin{table}[h]
\centering
\begin{tabular}{lcccc}
\toprule
Working point ($r_{\rm vtx}=30$~cm) & $\epsilon(h\to b\bar b)$ &  $1/\epsilon_{1b}$ &$1/\epsilon_{1c}$ & $\epsilon(g\to b\bar b)$ \\
\midrule $\geq 2$ systems                  & 0.91 & 11  & 17   & 0.87 \\
$m_{\rm vtx,2}>1.5$~GeV           & 0.69 & 25  & 104   & 0.67 \\
$m_{\rm vtx,2}>2.0$~GeV           & 0.51 & 57  & 950   & 0.50 \\
\bottomrule
\end{tabular}
\caption{
Signal tagging efficiency and background rejection factors, from $D$-decays, for jets containing a single $b$ or $c$ quark in the parton shower, assuming $r_{\rm vtx}=30$~cm.
Only jets that satisfied the $m_{\text{jet}}$ cut are included.
The last column is the tagging efficiency for $b$-jets which have a secondary $b\bar b$ pair from a hard gluon splitting in the parton shower.
}
\label{tab:wps}
\end{table}

The physically meaningful currency for our sensitivity study is not the tagging efficiency per reconstructed jet but per \emph{truth object}: the probability that a given $h\to b\bar b$ decay, or a given hard $b$ or $c$ quark, ends up tagged.
To calculate this, truth objects are matched to their highest-energy jet within $\Delta R < 0.5$ of the $h$ or $b$/$c$ momentum vector.
The total per-object tagging probabilities are then given by the product
\begin{align}
P_h \;\equiv&\; P(\text{hard jet}|h) \times P(\text{mass window}|h) \times
P(\text{flavor tag}|h),\\
P_b \;\equiv&\; P(\text{hard jet}|b) \times P(\text{mass window}|b) \times
P(\text{flavor tag}|b),\\
P_c \;\equiv&\; P(\text{hard jet}|c) \times P(\text{mass window}|c) \times
P(\text{flavor tag}|c),
\end{align}
where $P(\text{hard jet}\,|\,h)=0.998$, $P(\text{hard jet}\,|\,b)=0.93$ and $P(\text{hard jet}\,|\,c)=0.94$ are the probabilities that the jet was matched correctly to the truth-level primary object.
Furthermore we find $P(\text{mass window}|h)=0.86$, $P(\text{mass window}|b)=0.17$ and $P(\text{mass window}|c)=0.17$. 
Taking for example $r_{\rm vtx}=30$~cm and $m_{\rm vtx,2}>1.5$~GeV, we find $P_h=0.59$, $P_b=1/96$ and $P_c=1/219$.
Table~\ref{tab:perobj} gives the per-object working points; the per-$b$ fake rate here now \emph{includes} the irreducible contamination from gluons splitting ($g\to b\bar b$) in the parton shower of the $b$ jet.
For our application in \cref{sec:singprod} we find it optimal to prioritize signal efficiency over signal purity, and therefore choose the aggressive working point, highlighted in bold in \cref{tab:perobj}. 

Finally, we noticed that the performance of the tagger can be substantially improved with a three parameter likelihood over $m_{\text{vtx,2}}$, and the $\Delta R$ distance and energy imbalance between the two displaced vertex systems. 
The latter in particular can further reduce the $g\to b\bar b$ contribution.
For low efficiencies and high signal purity, the improvement is about an order of magnitude.
However it is not very helpful if high signal efficiency is preferred, as is the case in our study.
For brevity, we therefore omit further discussion of these additional features.

\begin{table}[h]
\centering
\begin{tabular}{lccccccccc}
\toprule
& \multicolumn{3}{c}{$\geq 2$ systems}
& \multicolumn{3}{c}{$m_{\rm vtx,2}>1.5$~GeV}
& \multicolumn{3}{c}{$m_{\rm vtx,2}>2.0$~GeV} \\
$r_{\rm vtx}$ & $P_h$ & $1/P_b$ & $1/P_c$
              & $P_h$ & $1/P_b$ & $1/P_c$
              & $P_h$ & $1/P_b$ & $1/P_c$\\
\midrule
5~cm  & 0.30 & 85 &  87   & 0.20 & 183 &  287    & 0.13 & 348 &    631  \\
10~cm & 0.54 & 66&   75   & 0.38 & 128 &  244    & 0.27 & 224 &  543    \\
30~cm & \textbf{0.79} & \textbf{52} &  \textbf{69}    & 0.59 & 96 &  219     & 0.44 & 163 & 487    \\
\bottomrule
\end{tabular}
\caption{Per-truth-object tagging probabilities: efficiency per fiducial $h\to b\bar b$ decay and inverse fake rate per fiducial hard $b$ or $c$ quark, for three usable-vertex radii and three tagger working points.
The working point we used in \cref{sec:singprod} is highlighted in bold.}
\label{tab:perobj}
\end{table}

\subsection{Desirable detector specifications}
From this study we can derive some qualitative recommendations for the detector specification.
Unsurprisingly, because of the high boost the maximum radius at which a vertex can be reconstructed is very relevant (see right-hand panel of \cref{fig:toy}).
Its importance is further enhanced if the energy of the jets is increased.
For our baseline vertex resolution, we had assumed specifications that are $\mathcal{O}(1)$ more ambitious than the current state-of-the-art detectors.
This turned out not to be necessary, and even a slightly worse detector resolution than that of current detectors would be sufficient.
The effectiveness of the tagger was also not particularly sensitive to the $p_T$ threshold for the tracks, which is perhaps not surprising given the high energy of the jets.

\section{Further details on single production\label{app:figures}}

\subsubsection*{Fully merged events}

\begin{figure}
    \centering
    \includegraphics[width=0.9\linewidth]{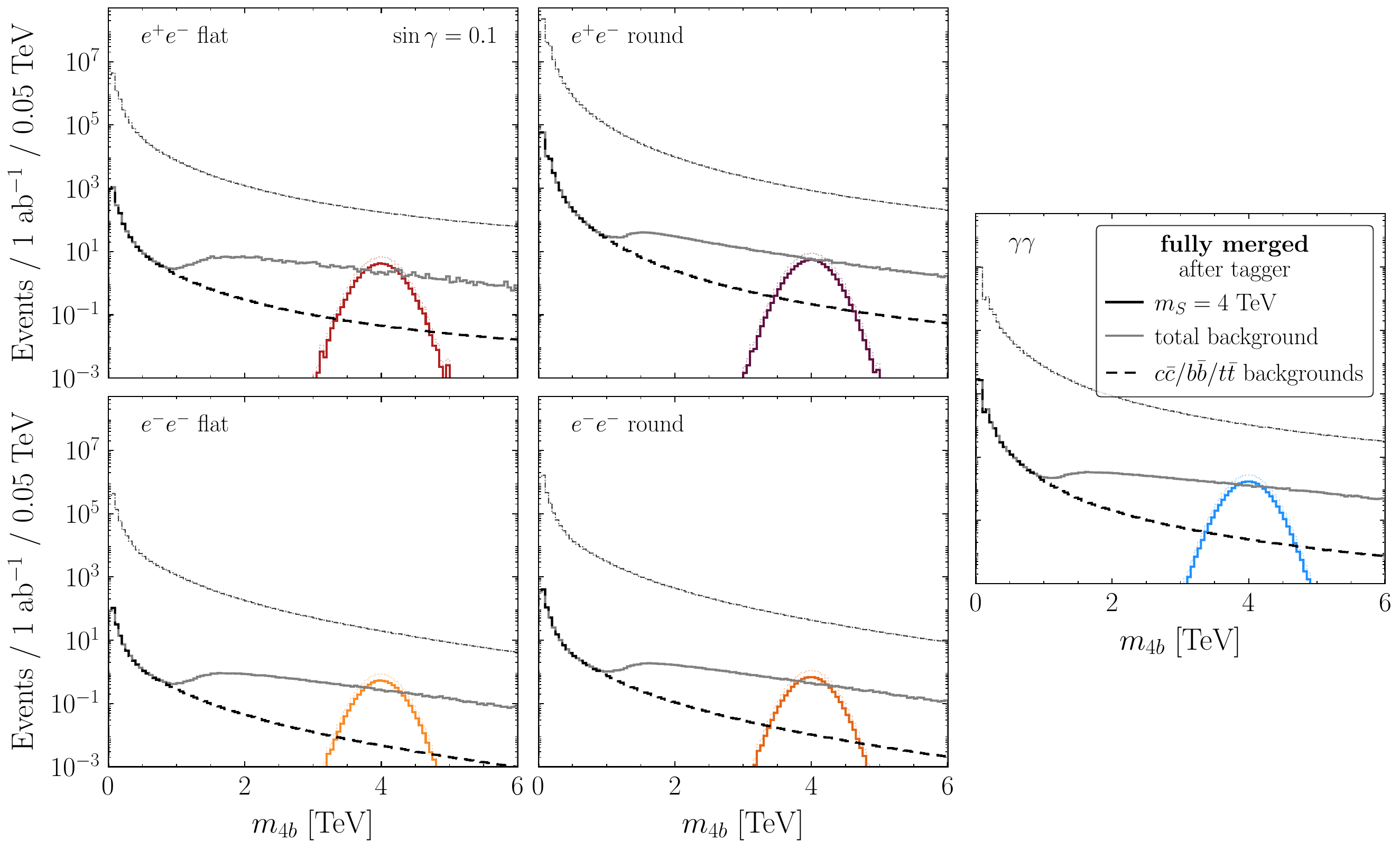}
    \caption{Invariant mass $m_{4b}$ distribution for signal with $m_S =\SI{4}{\TeV}$ and $\sin\gamma=0.1$ and background events in the fully merged category.
Faint lines show the total number of events in this category; bold lines show the number of events after the tagger is applied to both merged jets.
The combined contribution of $c\bar c$, $b\bar b$ and $t\bar t$ to the total background is shown in black.
    }
    \label{fig:inclusive_m4b_merged}
\end{figure}

\begin{figure}
    \centering
    \includegraphics[width=0.9\linewidth]{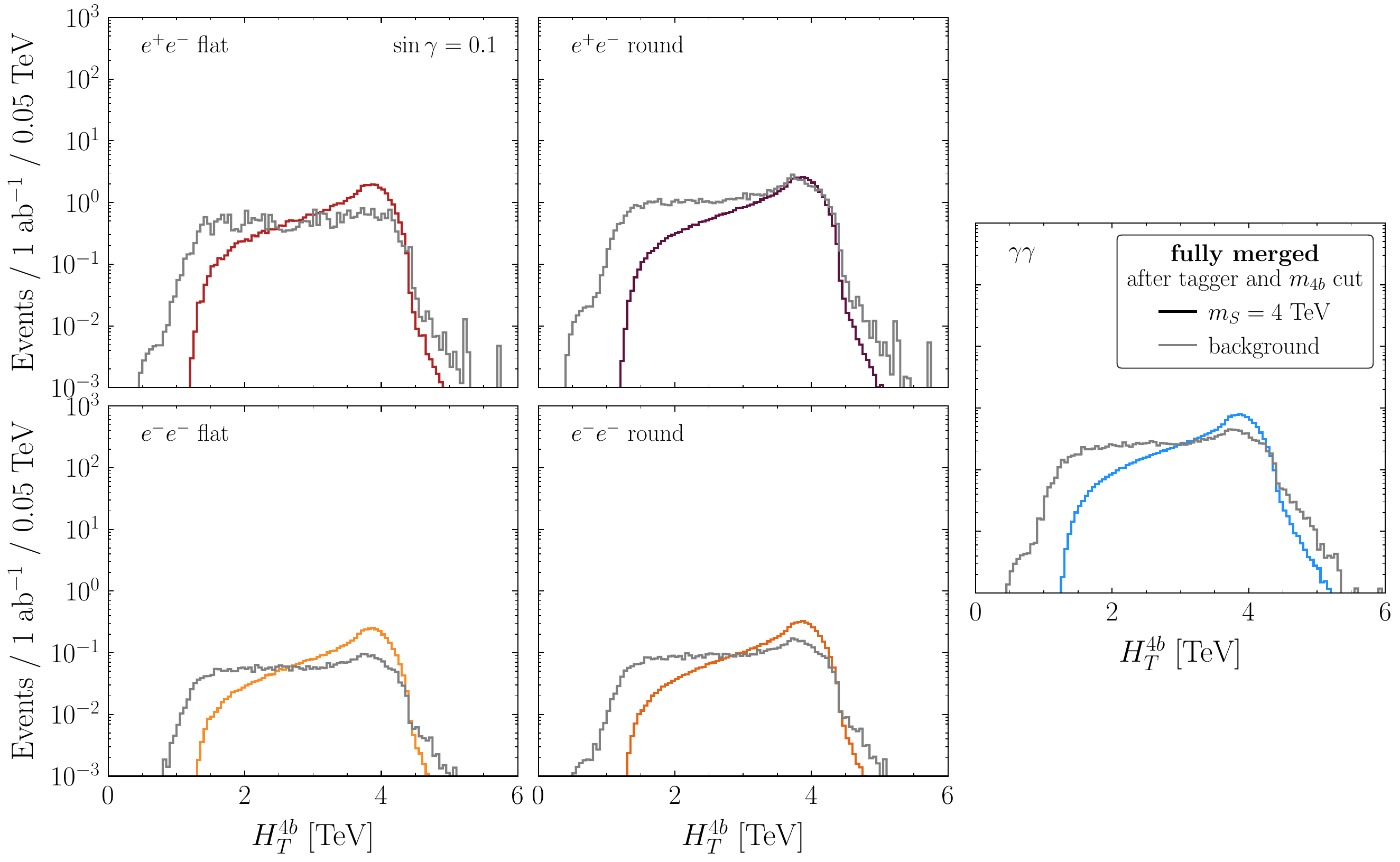}
    \caption{Scalar sum of transverse momenta of the $b$-jets $H_T^{4b}$ of signal ($m_S =\SI{4}{\TeV}$ and $\sin\gamma=0.1$) and background events in the fully merged category after the tagging requirements and the $m_{4b}$ cut. 
    }
    \label{fig:inclusive_HT4b_merged}
\end{figure}

\cref{fig:inclusive_m4b_merged} shows the invariant mass distribution $m_{4b}$ for the $m_S=\SI{4}{\TeV}$ signal and background after requiring both $b\bar b$ systems to be reconstructed as merged jets. The contribution of the fake double-$b$ jet backgrounds $c\bar{c}/b\bar{b}/t\bar{t}$ to the total background is shown as a black dashed line. The solid lines show clearly how the tagger efficiently suppresses this background compared to the faint lines before the tagger is applied. Because the fully merged selection favors boosted $h$ and $Z$ bosons, the low $m_{4b}$ region before tagging is dominated by the $c\bar{c}/b\bar b/t\bar{t}$ backgrounds. Suppression of this component by the tagger produces the dip in the background distribution, while the $hh$ background with sufficiently boosted Higgs bosons dominates at higher $m_{4b}$. For the signal, we see the expected resonance structure at $m_S=\SI{4}{\TeV}$.
\cref{fig:inclusive_HT4b_merged} depicts the total transverse energy distribution $H_T^{4b}$ for the fully merged category after the tagging requirements and $m_{4b}$ cut have been applied.

\subsubsection*{Partially merged events}

\begin{figure}
    \centering
    \includegraphics[width=0.9\linewidth]{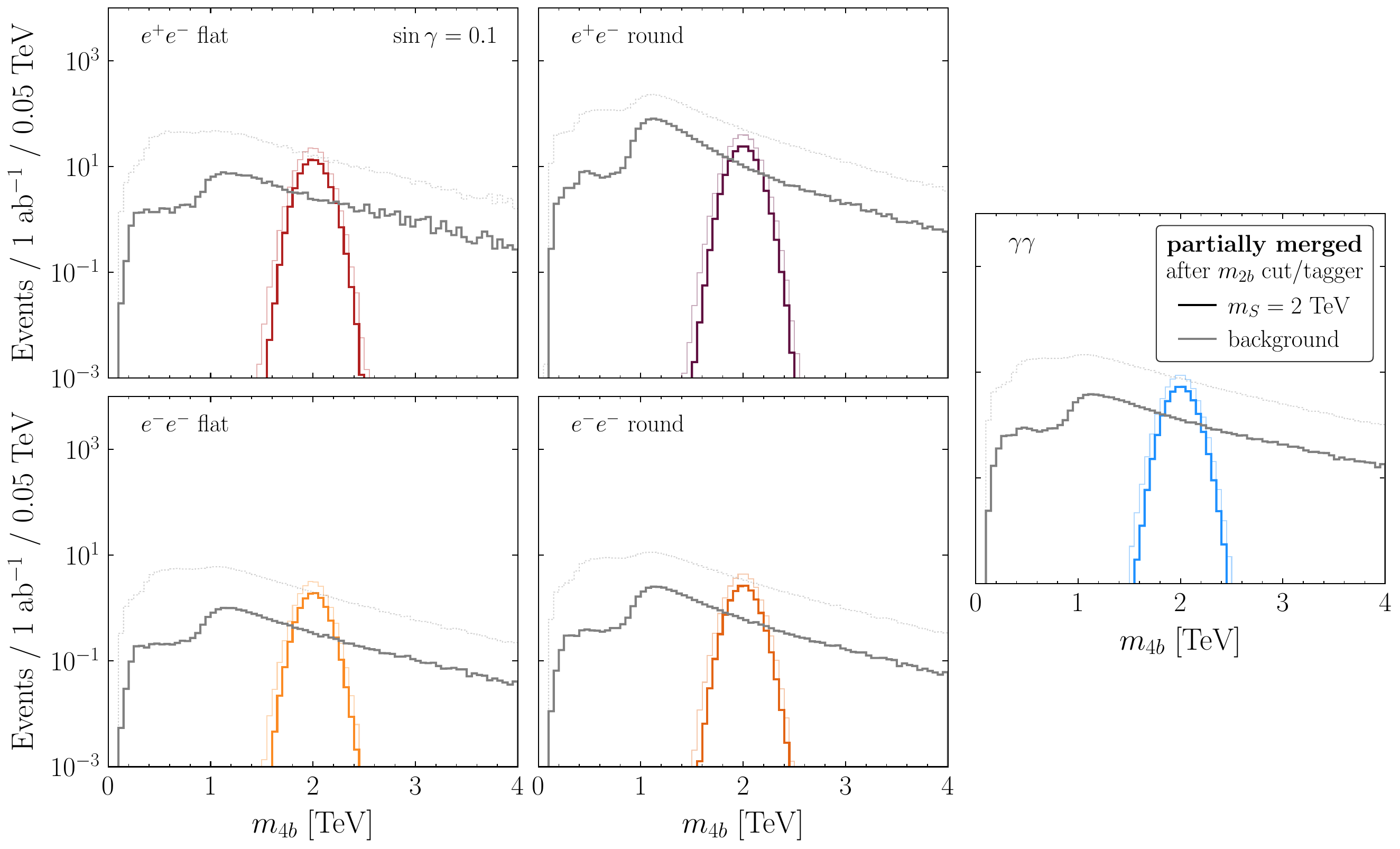}
    \caption{Invariant mass $m_{4b}$ distribution  for the signal with $m_S =\SI{2}{\TeV}$ and $\sin\gamma=0.1$ and background events in the partially merged category. Faint lines show the total number of events in this category, bold lines show the number of events after applying the merged-jet tagger and the $b$-tagging and Higgs-mass-window requirements to the resolved jet pair.}
    \label{fig:inclusive_m4b_semi}
\end{figure}

\begin{figure}
    \centering
    \includegraphics[width=0.9\linewidth]{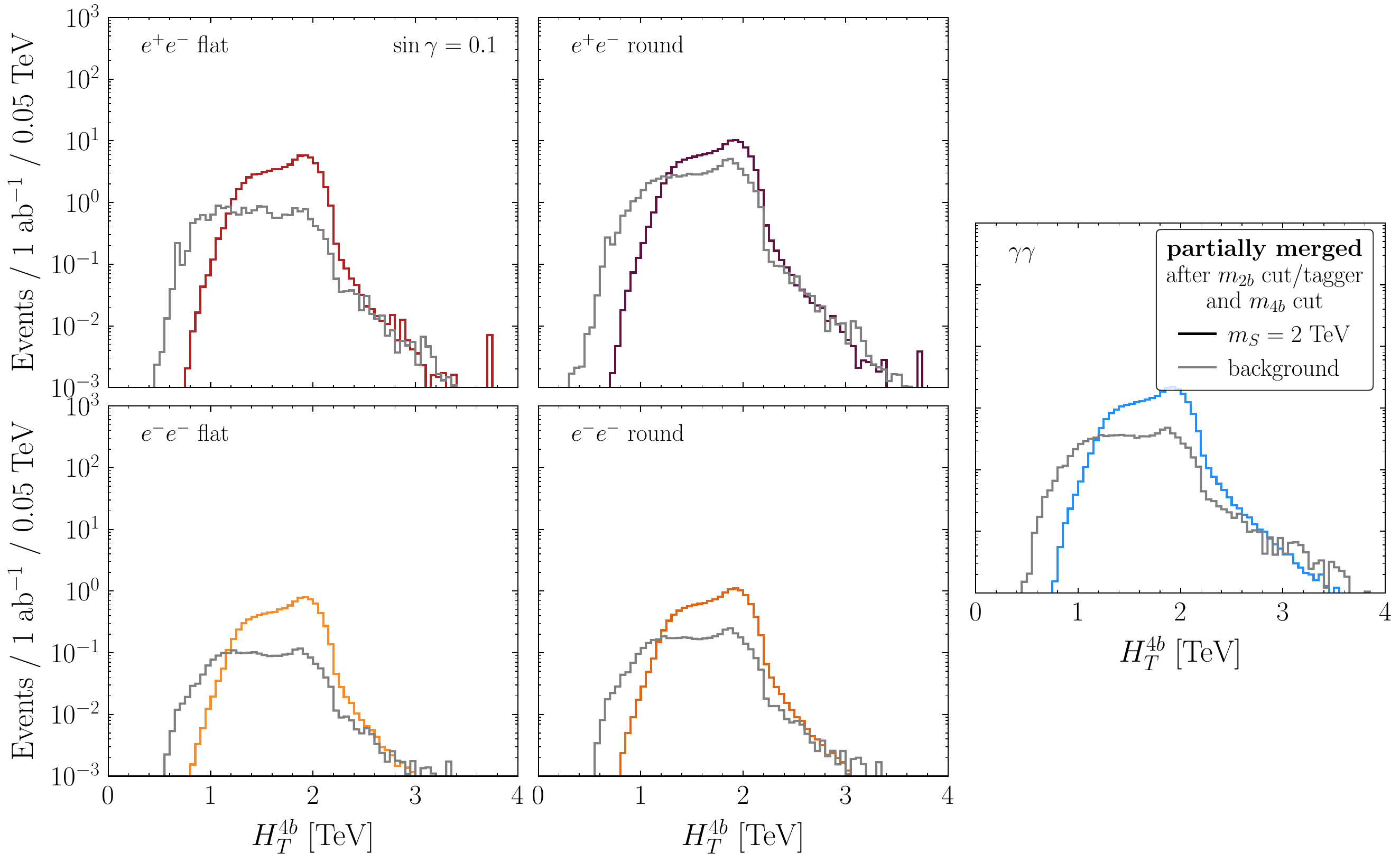}
    \caption{Scalar sum of the transverse momenta of the $b$-jets $H_T^{4b}$ of signal ($m_S =\SI{2}{\TeV}$ and $\sin\gamma=0.1$) and background events in the partially merged category after appropriate selections on $b$-jet pairs and on $m_{4b}$. 
    }
    \label{fig:inclusive_HT4b_semi}
\end{figure}

\cref{fig:inclusive_m4b_semi} shows the invariant mass distribution $m_{4b}$ for the $m_S=\SI{2}{\TeV}$ signal and background events in which one $b$-jet pair is merged and the other is resolved.
Faint lines show the total number of events before cuts, while bold lines show events after applying the $b$-tagging and Higgs-mass-window ($|m_{bb}^{(i)} - m_h| < 15\,\text{GeV}$) requirements to the resolved pair and the tagger efficiency to the merged jet.
Again, the resonance structure at $m_S=\SI{2}{\TeV}$ is recovered.
Compared to the fully separated category, fewer background events fall into the low $m_{4b}$ region as this category requires a boosted $h$ or $Z$ which will typically also lead to larger $m_{4b}$.
This effect is shifted to slightly lower $m_{4b}$ compared to the same effect in the fully merged category, where both parents of the $b$-jet pairs need to be boosted.
In~\cref{fig:inclusive_HT4b_semi} the total transverse energy distribution $H_T^{4b}$ after appropriate cuts on $m_{4b}$ and the $b$-jet pairs for the partially merged category is displayed, where the tagging efficiency is applied to the merged jet, while the resolved pair is subject to the $b$-tagging efficiency and the Higgs mass window. 

\clearpage

\section{Further details on pair production\label{app:pairprod}}

\subsection{The 4 $b$-jets + $4$ jets channel \label{app:4b4jets}}

For completeness, we present the cumulative signal efficiencies for the cuts presented in \cref{sec:singletpairprod}.
The ``in det.'' label refers to the $|\eta|<3$ rapidity cut on all decay products of the $S$, ``$p_T^j$'' is the $p_T$ cut on each jet.
The ``$m_{jj}$'' label refers to the requirement that each pair of jets ($b$-jets) must reconstruct an \ac{EW} boson ($h$).
The ``$m_{4j}$'' is the requirement that there exists a pairing of those objects such that both reconstruct $m_S$.
The remaining columns indicate the probability of finding a fixed number of merged di-jets from $W$, $Z$ or $h$ decay.
The $p^{tot}_T\equiv\big|\sum_j \vec{p}_{T,j}\big|$ label refers to the cut on the vector sum of the transverse momenta of all 8 jets. The selections are summarized in \cref{tab:cuts_double}, and the resulting cumulative efficiencies are shown in \cref{tab:cutflow_eff}.

\begin{table}[h]
  \centering
  
  \begin{tabular}{cl}
    \toprule
    Variable & Requirement \\
    \midrule
     $|\eta^j|$  &  $\leq 3$ \\
    $p_T^j $ & $> 30$~GeV \\
    $\Delta R_{jj}$         & $ \geq 0.4$ \\
      $|m_{2j}/m_{\text{EW}} - 1| $ & $\leq 0.12$ \\
    $|m_{4j}/m_S - 1|$          & $ \leq 0.1$ \\
    $\big|\sum_j \vec{p}_{T,j}\big|$ & $\geq 50$~GeV\\
    Number of merged jet pairs  ($\Delta R_{jj}$< 0.4)      & $\leq 2$\\
    \bottomrule
  \end{tabular}
    \caption{Event selection cuts applied to the $SS \to 4b + 4j$ signal.\label{tab:cuts_double}}
\end{table}

\newcommand{\sci}[2]{\ensuremath{#1\cdot 10^{#2}}}

\begin{table}[h]
\centering
\small
\setlength{\tabcolsep}{3.5pt}
\begin{tabular}{c ccccc ccccc}
\toprule
 & \multicolumn{5}{c}{Cumulative selection} & \multicolumn{5}{c}{Number of merged jet-pairs ($\Delta R \le 0.4$)} \\
\cmidrule(lr){2-6} \cmidrule(lr){7-11}
$m_S$\,[TeV] & in det. & $p_T^j$ & $m_{4j}$ & $m_{jj}$ & $p^{tot}_T$
            & $0$ & $1$ & $2$ & $3$ & $4$ \\
\midrule
0.5 & 0.86 & 0.36 & 0.32 & 0.24 & 0.21 & \sci{1.6}{-1} & \sci{3.0}{-2} & \sci{1.2}{-2} & \sci{5.2}{-3} & \sci{2.5}{-3} \\
1.0 & 0.92 & 0.53 & 0.46 & 0.34 & 0.31 & \sci{1.2}{-1} & \sci{9.5}{-2} & \sci{7.7}{-2} & \sci{1.3}{-2} & \sci{2.4}{-3} \\
1.5 & 0.94 & 0.63 & 0.54 & 0.40 & 0.37 & \sci{2.6}{-2} & \sci{9.8}{-2} & \sci{1.7}{-1} & \sci{6.4}{-2} & \sci{9.1}{-3} \\
2.0 & 0.95 & 0.70 & 0.60 & 0.45 & 0.41 & \sci{5.9}{-3} & \sci{4.2}{-2} & \sci{1.6}{-1} & \sci{1.6}{-1} & \sci{5.4}{-2} \\
2.5 & 0.96 & 0.75 & 0.64 & 0.47 & 0.44 & \sci{1.5}{-3} & \sci{1.7}{-2} & \sci{9.9}{-2} & \sci{1.9}{-1} & \sci{1.3}{-1} \\
3.0 & 0.96 & 0.78 & 0.66 & 0.49 & 0.46 & \sci{9.7}{-4} & \sci{7.9}{-3} & \sci{6.0}{-2} & \sci{1.8}{-1} & \sci{2.1}{-1} \\
3.5 & 0.96 & 0.80 & 0.67 & 0.50 & 0.48 & \sci{3.4}{-4} & \sci{4.2}{-3} & \sci{3.9}{-2} & \sci{1.5}{-1} & \sci{2.8}{-1} \\
4.0 & 0.96 & 0.83 & 0.70 & 0.52 & 0.49 & \sci{2.0}{-4} & \sci{2.7}{-3} & \sci{2.5}{-2} & \sci{1.3}{-1} & \sci{3.4}{-1} \\
4.5 & 0.97 & 0.84 & 0.71 & 0.53 & 0.50 & \sci{2.0}{-4} & \sci{2.0}{-3} & \sci{2.1}{-2} & \sci{9.3}{-2} & \sci{3.8}{-1} \\
\bottomrule
\end{tabular}
\caption{Cumulative cut-flow efficiencies for the double-production search. The last five columns are exclusive bins in the number of
jet-pairs failing $\Delta R > 0.4$, applied after all preceding cuts, and
therefore sum to the $p^{tot}_T$ column. }
\label{tab:cutflow_eff}
\end{table}

\subsection{8 jets channel\label{app:8jetsbackground}}
When all $WW$, $ZZ$ and $hh$ decay modes for each $S$ are accounted for, the signal rate for having 8 jets in the final state is about a factor of 3 larger than for the 4 jets + 4 $b$-jets final state we considered in \cref{sec:singletpairprod}.
We nevertheless found that the reach is slightly weaker, due to the larger $\gamma\gamma$ induced backgrounds.
We explain this below.

Following the discussion in \cref{sec:singletpairprod}, we impose a modest (50 GeV) cut on the vector sum of the transverse momenta of the 8 jets.
This implies that backgrounds for this signature are four vector bosons ($4W$ and $2W 2Z$) produced through \ac{VBF}, e.g.~$e^+e^-\to W^+W^-W^+W^-\nu_e\bar\nu_e$, or in association with a fifth \ac{EW} boson, e.g.~$\gamma\gamma \to W^+W^-W^+W^- Z$, where the fifth boson is needed to pass the aforementioned cut.
At collision energies approaching 10 TeV, these processes are subtle to calculate because of their large phase space and because fixed order electroweak perturbation theory is starting to converge slowly.
They are worthy of a careful, dedicated study which is beyond the scope of our present work.
We instead rely on a number of approximations to estimate these backgrounds at the $\mathcal{O}(1)$ level.
The background processes we considered in this analysis are
\begin{align}
e^+e^-&\to W^+W^-W^+W^-\nu_e\bar\nu_e\label{eq:multiWfirst}\\
e^+e^-&\to W^+W^-ZZ\nu_e\bar\nu_e\\
e^-e^-&\to W^+W^-W^+W^-e^-e^-\\
e^-e^-&\to W^+W^-ZZe^-e^- \\
e^+e^-&\to W^+W^-W^+W^-Z\label{eq:multiWlast}\\
\gamma\gamma&\to W^+W^-W^+W^-Z \label{eq:gg_to_4WZ}\\
\gamma\gamma&\to W^+W^-ZZZ \label{eq:gg_to_2W3Z}
\end{align}
The cross section of the $\gamma\gamma$-induced process in \eqref{eq:gg_to_4WZ} is by far the largest, followed by \eqref{eq:gg_to_2W3Z}, which is between 20\% and 40\% of the size of \eqref{eq:gg_to_4WZ}.
The processes in \eqref{eq:gg_to_4WZ} and \eqref{eq:gg_to_2W3Z} dominate because they arise at lower order in the weak coupling than the \ac{VBF} processes and have a $t$-channel enhancement that is absent for $e^+e^-\to W^+W^-W^+W^-Z$.

\sisetup{
  scientific-notation = true,
  round-mode = figures,
  round-precision = 2,
  exponent-product = \times,
}

\begin{table}[tbp]
\begin{tabular}{l S S S S}
\hline
{collider} & {$m_S=0.5$ TeV} & {$m_S=1$ TeV} & {$m_S=1.5$ TeV} & {$m_S=2$ TeV} \\
\hline
$e^+e^-$ round    & 8.258e3 & 7.798e3 & 5.207e3 & 2.940e3 \\
$e^+e^-$ flat     & 1.678e3 & 1.821e3 & 1.379e3 & 8.736e2 \\
$e^-e^-$ round    & 7.046e2 & 7.685e2 & 5.863e2 & 3.745e2 \\
$e^-e^-$ flat     & 3.279e2 & 3.627e2 & 2.794e2 & 1.797e2 \\
$\gamma\gamma$  & 2.506e3 & 2.961e3 & 2.541e3 & 2.000e3 \\
\hline
\end{tabular}

\vspace{0.1cm}

\begin{tabular}{l S S S}
\hline
{collider} & {$m_S=2.5$ TeV} & {$m_S=3$ TeV} & {$m_S=3.5$ TeV} \\
\hline
$e^+e^-$ round    & 1.606e3 & 8.592e2 & 3.840e2 \\
$e^+e^-$ flat     & 5.275e2 & 3.058e2 & 1.481e2 \\
$e^-e^-$ round    & 2.281e2 & 1.333e2 & 6.506e1 \\
$e^-e^-$ flat     & 1.099e2 & 6.442e1 & 3.151e1 \\
$\gamma\gamma$  & 1.557e3 & 1.210e3 & 8.678e2 \\
\hline
\end{tabular}

\caption{Estimated background yields for the dominant $\gamma\gamma\to W^+W^-W^+W^-Z$ process after cuts, for a geometric luminosity of $10\,\mathrm{ab}^{-1}$.}
\label{tab:backgrounddoubleS}
\end{table}

We estimate the $\gamma\gamma\to W^+W^-W^+W^-Z$ cross section by simulating the cross section for one fewer \ac{EW} boson with MadGraph and subsequently multiplying by an estimated penalty factor for the emission of an additional \ac{EW} boson.
Concretely, we make the approximation
\begin{equation}
\sigma(\gamma\gamma\to W^+W^-W^+W^-Z)\approx \sigma(\gamma\gamma\to W^+W^-W^+W^-)\times \frac{\sigma(\gamma\gamma\to W^+W^-W^+W^-)}{\sigma(\gamma\gamma\to W^+W^-Z)}\
\end{equation}
where the cross sections are the inclusive cross sections for these processes, but subject to a $|\eta|<3$ cut on all final states.
The latter is needed to eliminate parts of the phase space with large rates but that fall outside our assumed detector acceptance.
This should be considered an $\mathcal{O}(1)$-level estimate.
We conservatively assume that all such background events pass the 50 GeV cut on the transverse momentum of the four-boson system.
We then apply all other cuts described above to the $\gamma\gamma\to W^+W^-W^+W^-$ sample, to find the efficiency for each hypothesized value of $m_S$.
This is used to calculate the expected number of events shown in \cref{tab:backgrounddoubleS} by weighting the cross section $\times$ efficiency with the luminosity spectra for each collider.

The remaining processes in \eqref{eq:multiWfirst} to \eqref{eq:multiWlast} were also simulated with MadGraph, using the effective vector approximation \cite{Dawson:1984gx,Kane:1984bb,Ruiz:2021tdt} for the \ac{VBF} topologies. 
After applying the kinematic selection cuts described in \cref{sec:pairprompt} and luminosity weighting, their corresponding event yield is between one and two orders of magnitude smaller than that for $\gamma\gamma\to W^+W^-W^+W^-Z$, across all colliders.
We therefore neglect them in \cref{tab:backgrounddoubleS}.
Finally, a priori three boson processes, such as $WWZ$ production, can supply a background if the decay products of one boson each fake an ``\ac{EW} boson jet''.
For a $10\%$ mistag rate, we verified that this class of backgrounds can also be neglected.
After accounting for these backgrounds, we find that the reach is at best comparable to the 4 jets + 4 $b$-jets channel, and in most cases slightly worse.
In a full analysis at a future experiment one could of course improve the sensitivity by statistically combining all decay channels, which we do not attempt here.
We emphasize that the values in \cref{tab:backgrounddoubleS} should be taken as order-of-magnitude estimates only.
We merely use them to justify our choice to focus on the 4 jets + 4 $b$-jets channel over the 8 jets channel, and a future more detailed study would be desirable.

\subsection{Displaced analysis}

\begin{figure}
    \centering
    \includegraphics[width=1\linewidth]{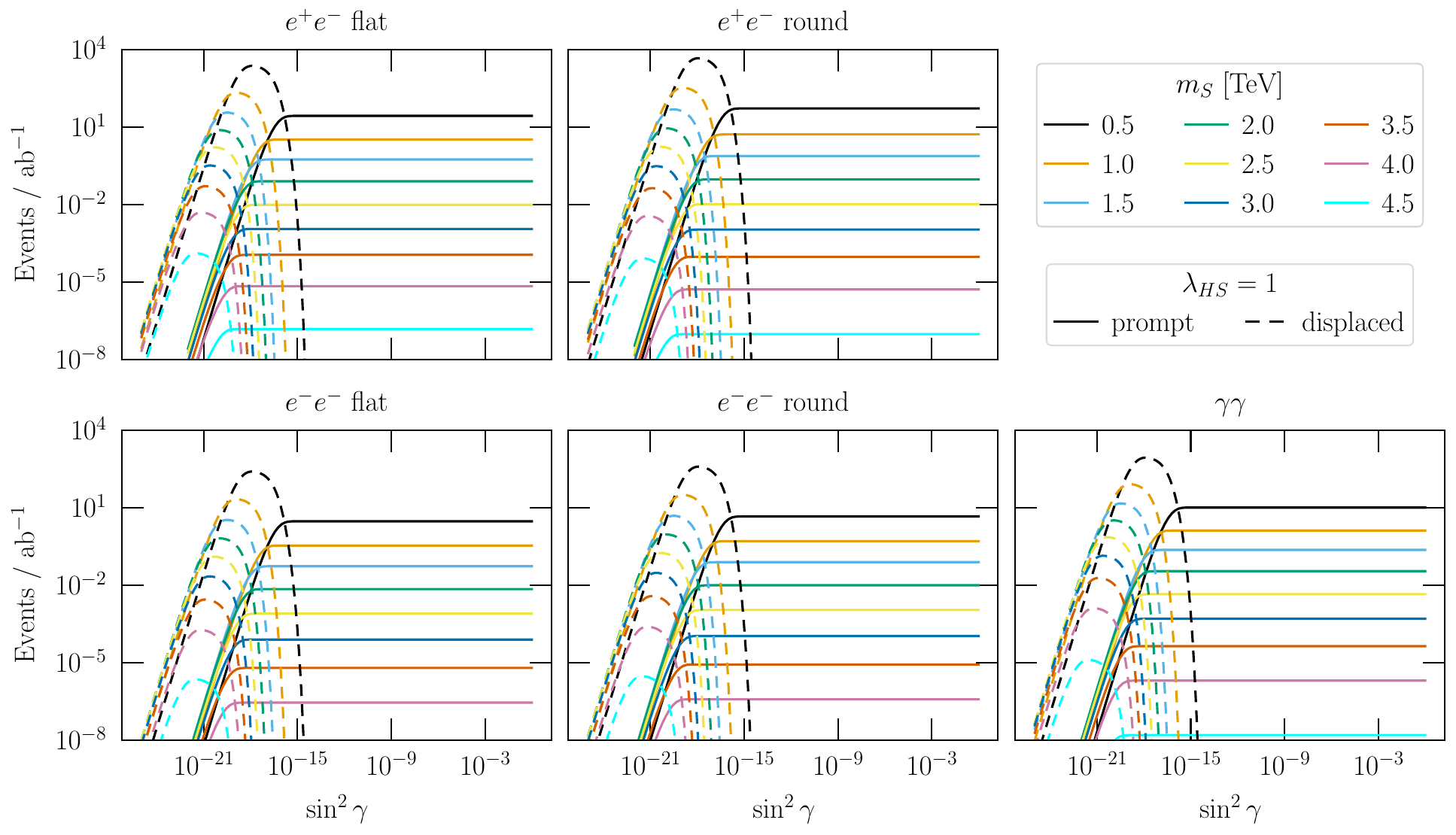}
    \caption{Number of events from pair-produced scalars passing the selections for the prompt (solid) and displaced (dashed) searches as a function of $\sin^2\gamma$ for different scalar masses.}
    \label{fig:double_prompt_displaced_s2g}
\end{figure}

The expected number of events per $\lambda_{HS}^2$ per $\rm {ab}^{-1}$ is shown in Fig.~\ref{fig:double_prompt_displaced_s2g} as a function of $\sin^2\gamma$ for different scalar masses. 
The plot also shows the number of events expected to pass the prompt search selections, weighted by the probability for both scalars to decay within $R_T\leq 5$ cm from the interaction point.

\FloatBarrier
\bibliographystyle{JHEP}
\bibliography{bibliography}
\end{document}